\documentclass[sigplan, 10pt, screen, authorversion]{acmart}

\acmSubmissionID{4}

\usepackage{times}
\usepackage{graphicx}
\usepackage{listings, listings-rust}
\usepackage{xcolor}
\usepackage{float}
\usepackage{fancyvrb}
\usepackage{url}
\usepackage{todonotes}
\usepackage{booktabs}
\usepackage{subfigure}
\usepackage{amsmath}
\usepackage{caption}
\usepackage{tikz}
\usepackage{ifdraft}

\usepackage{xcolor}
\usepackage{braket}   
\usepackage{amsfonts} 
\usepackage{multicol}

\usepackage[english]{babel}
\usepackage[utf8]{inputenc}
\usepackage{algorithm}
\usepackage{algpseudocode}

\usetikzlibrary{tikzmark}
\tikzset{mycircled/.style={circle,draw,inner sep=0.1em,line width=0.04em}}
\def\rcwhite#1{\tikzmarknode[mycircled,draw=red,fill=red]{t1}{\textcolor{white}{#1}}}
\def\wcblack#1{\tikzmarknode[mycircled,draw=black,fill=white]{t1}{\textcolor{black}{#1}}}

\makeatletter
\DeclareRobustCommand\ttfamily
{\not@math@alphabet\ttfamily\mathtt
	\fontfamily\ttdefault\small\selectfont}
\makeatother

\newcounter{ecount}

\def\BEGINITEMIZE{\begin{itemize}}
\def\BEGINENUMERATE{\begin{enumerate}}
\def\ENDITEMIZE{\end{itemize}}
\def\ENDENUMERATE{\end{enumerate}}

\makeatletter
\long\def\unmarkedfootnote#1{{\long\def\@makefntext##1{##1}\footnotetext{#1}}}
\makeatother

\begin{document}

\acmYear{2022}\copyrightyear{2022}
\setcopyright{acmlicensed}
\acmConference[EuroSys '22]{Seventeenth European Conference on Computer Systems}{April 5--8, 2022}{RENNES, France}
\acmBooktitle{Seventeenth European Conference on Computer Systems (EuroSys '22), April 5--8, 2022, RENNES, France}
\acmPrice{15.00}
\acmDOI{10.1145/3492321.3519553}
\acmISBN{978-1-4503-9162-7/22/04}



\title{Isolating Functions at the Hardware Limit with Virtines}

\author{Nicholas C. Wanninger}
\email{ncw@u.northwestern.edu}
\affiliation{
    \institution{Northwestern University}
    \streetaddress{633 Clark St.}
    \city{Evanston}
    \state{Illinois}
    \country{USA}
    \postcode{60208}
}
\authornote{A majority of this work was done while at Illinois Institute of Technology.}

\author{Joshua J. Bowden}
\email{jbowden@hawk.iit.edu}
\affiliation{
    \institution{Illinois Institute of Technology}
    \streetaddress{10 W. 31st St.}
    \city{Chicago}
    \state{Illinois}
    \country{USA}
    \postcode{60616}
}

\author{Kirtankumar Shetty}
\email{kshetty11@hawk.iit.edu}
\affiliation{
    \institution{Illinois Institute of Technology}
    \streetaddress{10 W. 31st St.}
    \city{Chicago}
    \state{Illinois}
    \country{USA}
    \postcode{60616}
}

\author{Ayush Garg}
\email{agarg22@hawk.iit.edu}
\affiliation{
    \institution{Illinois Institute of Technology}
    \streetaddress{10 W. 31st St.}
    \city{Chicago}
    \state{Illinois}
    \country{USA}
    \postcode{60616}
}

\author{Kyle C. Hale}
\email{khale@cs.iit.edu}
\affiliation{
    \institution{Illinois Institute of Technology}
    \streetaddress{10 W. 31st St.}
    \city{Chicago}
    \state{Illinois}
    \country{USA}
    \postcode{60616}
}

\begin{abstract}

An important class of applications, including programs that leverage third-party libraries,
programs that use user-defined functions in databases, and serverless applications, 
benefit from isolating the execution of
untrusted code at the granularity of individual functions or function
invocations.  However, existing isolation mechanisms were not designed for this
use case; rather, they have been adapted to it.  We introduce \textit{virtines},
a new abstraction designed specifically for function granularity isolation, and
describe how we build virtines from the ground up by pushing hardware
virtualization to its limits.  Virtines give developers fine-grained control in
deciding which functions should run in isolated environments, and which should
not. The virtine abstraction is a general one, and we demonstrate a prototype
that adds extensions to the C language. We present a detailed analysis of the
overheads of running individual functions in isolated VMs, and guided
by those findings, we present Wasp, an embeddable hypervisor that allows
programmers to easily use virtines. We describe several representative scenarios
that employ individual function isolation, and demonstrate that virtines can be
applied in these scenarios with only a few lines of changes to existing
codebases and with acceptable slowdowns.

\end{abstract}

\settopmatter{printfolios=false}
\maketitle

\renewcommand{\shortauthors}{N. Wanninger, J. Bowden, K. Shetty, A. Garg, and K. Hale}

\section{Introduction}
\label{sec:intro}

Vulnerabilities in critical applications can lead to information leakage, data
corruption, control-flow hijacking, and other malicious activity. If vulnerable
applications run with elevated privileges, the entire system may be
compromised~\cite{CVE-2021-3156, CVE-2014-0160}. Systems that execute code from
untrusted sources must then employ isolation mechanisms to ensure data secrecy,
data integrity, and execution integrity for critical software
infrastructure~\cite{DAUTENHAHN:2015:NESTEDKERNEL, NARAYANAN:2019:LXD,
SWIFT:2002:NOOKS, HSU:2016:SMV, HEDAYATI:2019:HODOR, GHOSN:2021:ENCLOSURES,
MCCUNE:2010:TRUSTVISOR, LIU:2015:SECAGE, BITTAU:2008:WEDGE,
HALE:2014:GUARDMODS, VAHLDIEK:2019:ERIM, MCCUNE:2008:FLICKER}. This isolation
typically happens in a coarse-grained fashion, but an important class of
applications require isolation at the granularity of individual functions or
distinct invocations of such functions. Long-standing examples include the use
of untrusted library functions by critical applications and user-defined
functions (UDFs) in databases, while serverless functions represent an
important emerging example. Existing isolation mechanisms, however, were not
designed for individual functions. Applications that leverage this isolation
model must resort to repurposing off-the-shelf mechanisms with mismatched
design goals to suit their needs. For example, databases limit UDFs to run in a
managed language like Java or Javascript~\cite{MICHAEL:1998:UDF, ORACLE-UDFS},
and serverless platforms repurpose containers to isolate users' stateless
function invocations from one another. The latter example is particularly
salient today. As others have shown at this venue, formidable challenges
(particularly cold-start latency) arise when using containers for individual
function execution~\cite{CADDEN:2020:SEUSS}. These challenges stem from
contorting the container abstraction to fit an unintended usage model.

Guided by these examples, we introduce \textit{virtines}, a new abstraction
\textit{designed} for isolating execution at function call granularity using
hardware virtualization. Data touched by a virtine is automatically encapsulated
in the virtine's isolated execution environment. This environment implements an
abstract machine model that is not constrained by the traditional x86 platform.
Virtines can seamlessly interact with the host environment through a checked
hypervisor interposition layer. With virtines, programmers annotate critical
functions in their code using language extensions, with the semantics that a
single virtine will run in its own, isolated virtual machine environment. While
virtines require code changes, these changes are minimal and easy to
understand. Our current language extensions are for C, but we believe they can
be adapted to most languages.

The execution environments for virtines (including parts of the hypervisor) are
tailored to the code inside the isolated functions; a virtine image contains
\textit{only} the software that a function needs. We present a detailed,
ground-up analysis of the start-up costs for virtine execution environments,
and apply our findings to construct small and efficient virtine images.
Virtines can achieve isolated execution microsecond scale startup latencies and
limited slow-down relative to native execution. They are supported by a custom,
user-space runtime system implemented using hardware virtualization called
Wasp, which comprises a small, embeddable hypervisor that runs on both Linux
and Windows. The Wasp runtime provides mechanisms to enforce strong virtine
isolation by default, but isolation policies can be customized by users.

Our contributions in this paper are as follows:

\BEGINITEMIZE

    \item We introduce {\em virtines}, programmer-guided abstractions that allow
        individual functions to run in light-weight, virtualized execution
        environments.

    \item We present a prototype embeddable hypervisor framework, Wasp, that
        implements the virtine abstraction. Wasp runs as a Type-II micro-hypervisor on both Linux
        and Windows.

    \item We provide language extensions for programming with virtines in C that
        are conceptually simple.

    \item We evaluate Wasp's performance using extensive microbenchmarking, and
        perform a detailed study of the costs of virtine execution environments.

    \item We demonstrate that it requires minimal effort to incorporate
        virtines into software components used in 
        representative scenarios involving function isolation: namely,
        untrusted or sensitive library functions (OpenSSL) and managed language runtimes
        (Javascript).  The virtine versions incur acceptable
        slow-down while using strong hardware isolation.

\ENDITEMIZE

\section{Virtines}
\label{sec:virtines}

A {\em virtine} provides an isolated execution environment using lightweight
virtualization. Virtines consist of three components: a toolchain-generated
binary to run in a virtual context, a hypervisor that facilitates the VM's only
external access (Wasp), and a host program which specifies virtine isolation
policies and drives Wasp to create virtines. When invoked, virtines run
synchronously from the caller's perspective, leading them to appear and act
like a regular function invocation. However, virtines could, given support in
the hypervisor, behave like asynchronous functions or futures.\footnote{For
example, like \textit{goroutines}, as in Gotee~\cite{GHOSN:2019:SECROUTINES}:
\url{https://gobyexample.com/goroutines}} As with most code written to execute
in a different environment from the host, such as CUDA code~\cite{CUDA_PROG_GUIDE} or
SGX enclaves~\cite{INTEL-SGX}, there are constraints on what virtine code can
and cannot do. Due to their isolated nature, virtines have no direct access to
the caller's environment (global variables, heap, etc.). A virtine can, however,
accept arguments and produce return values like any normal function.
These arguments and return values are marshalled automatically by the virtine
compiler when using our language extensions.\footnote{When using the virtine
runtime library directly, developers must currently marshal arguments and
return values manually, though we are currently developing an IDL to ease this
process (like SGX's EDL~\cite{INTEL_SGX_SDK}).}

\begin{figure}[t]
    \centering
    \includegraphics[width=0.8\columnwidth]{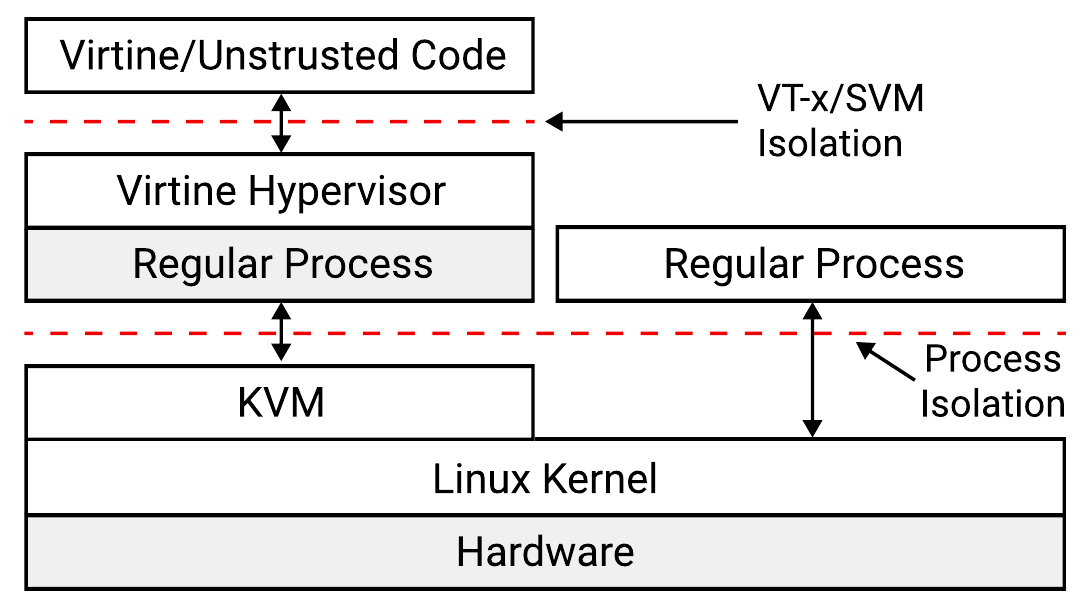}
    \caption{How virtines fit into the software stack.}
    \label{fig:virtine-overview}
\end{figure}

Unlike traditional hypervisors, a virtine hypervisor need not--and we suspect
in most cases will not--emulate every part of the x86 platform, such as PCI,
ACPI, interrupts, or legacy I/O. A virtine hypervisor therefore implements an
abstract machine model designed for and restricted to the intentions of the
virtine. Figure~\ref{fig:virtine-overview} outlines the architecture and data
access capabilities (indicated by the arrows) of a virtine compared to a
traditional process abstraction. A host program that uses (links against) the
embeddable virtine hypervisor has some, but not necessarily all, of its
functions run as virtines. We refer to such a host process as a virtine
\textit{client}. If the virtine context wishes to access any data or service
outside of its isolated environment, it must first request access from the
client via the hypervisor. Virtines exist in a default-deny environment, so
the hypervisor must interpose on all such requests. While the hypervisor
provides the interposition mechanism~\cite{BAUMANN:2013:BASCULE}, the virtine client has the option to
implement a hypercall policy, which determines whether or not an individual
request will be serviced. The capabilities of a virtine are determined by (1)
the hypervisor, (2) the runtime within a virtine image, and (3) policies
determined by the virtine client.

Virtines are constructed from a subset of an application's call graph.
Currently, the decision where the ``cut'' in the call graph is made by the
programmer, but making this choice automatically in the compiler is
possible~\cite{LIND:2017:GLAMDRING, LIU:2015:SECAGE}. Since a virtine constitutes only
a subset of the call graph, virtine images are typically small
($\sim$16KB), and are statically compiled binaries containing all required
software. Shared libraries violate our isolation requirements, as we will see in
Section~\ref{sec:objectives}.

While the runtime environment that underlies a function running in virtine context can vary, we expect
that in most cases this environment will comprise a limited, kernel-mode only, software layer. This may
mean no scheduler, virtual memory, processes, threads, file systems, or any
other high-level constructs that typically come with running a fully-featured
VM. This is not, however, a requirement, and virtines can take advantage of
hardware features like virtual memory, which can lead to interesting
optimizations like those in Dune~\cite{BELAY:2012:DUNE}. Additional
functionality {\em must} be provided by adding the functionality to the virtine
environment or by borrowing functionality from the hypervisor. Adding this
functionality should be done with care, as interactions with the hypervisor
come with costs, both in terms of performance and isolation. In this paper, we
provide two pre-built virtine execution environments
(Section~\ref{sec:exec-environ}), but we envision a rich virtine
ecosystem could develop from which an execution environment could be selected.
These environments could also be
synthesized automatically. Note that one possible execution environment for a
virtine is a unikernel. However, unikernels are typically designed with a
standard ABI in mind (e.g., binary compatible with Linux). Virtine execution environments are instead co-designed with
the virtine client, and allow for a wide variety of virtual platforms which may support
non-standard ABIs.

\section{Design}
\label{sec:design}

In this section, we describe our isolation and safety objectives in developing the
virtine abstraction. We then discuss how to achieve these goals using hardware
and software mechanisms.

\subsection{Safety Objectives}
\label{sec:objectives}


\paragraph*{Host execution and data integrity}
Host code and data cannot be modified, and its control flow cannot be hijacked
by a virtine running untrusted or adversarial code.

\paragraph*{Virtine execution and data integrity}
The private state of a virtine must not be affected by another virtine running
untrusted or adversarial code. Thus, data secrecy must be maintained between
virtines.

\paragraph*{Virtine isolation}
Host data secrecy must also be maintained, so virtines may not interact with
any data or services outside of their own address space other than what is
explicitly permitted by the virtine client's policies.

These objectives are similar to sandboxing in web browsers, where some
components (tabs, extensions) within the same address space are untrusted,
meaning that intra-application interactions must cross isolation boundaries. In
Google Chrome, for example, process isolation and traditional security
restrictions are used to achieve this isolation~\cite{LIU-12-CHROME-EXTENSIONS,
CHROME-SANDBOX-LINUX}. Unfortunately, as with most software, bugs have allowed
attackers to access user data, execute arbitrary code, or just crash the
browser with carefully crafted JavaScript~\cite{CVE-2009-2555,CVE-2018-6056,
CVE-2018-18342, CVE-2009-2935,CVE-2017-2505}.

\subsection{Threat Model}
\label{sec:threat}

Code that runs in virtine context can still suffer from software bugs such as
buffer overflow vulnerabilities. We therefore assume an adversarial model,
where attacks that arise from such bugs may occur, and where a virtine can
behave maliciously. We assume the hypervisor (Wasp) and the host kernel are
trusted, similar to prior work~\cite{ALPERNAS:2018:SECURE-SERVERLESS}. In
addition, we assume that the virtine client---in particular, its hypercall
handlers---are trusted and implemented correctly. These handlers must take care
to assume that inputs have \textit{not} been properly sanitized, and may even
be intentionally manipulated. Along with using best practices, we assume
hypercall handlers are careful when accessing the resources mapped to a
virtine, for example checking memory bounds before accessing virtine memory,
validating potentially unsafe arguments, and correctly following the access
model that the virtine requires. We assume that virtines do not share state
with each other via shared mappings, and that they cannot directly access host
memory. Additionally, we assume that microarchitectural and host kernel
mitigations are sufficient to eliminate side channel attacks. Note that we do
not expect end users to implement their own virtine clients. We instead assume
that runtime experts will develop the virtine clients (and corresponding hypercall
handlers). In this sense, our assumptions about the virtine client's integrity
are similar to those made in cloud platforms that employ a user-space device
model (e.g., QEMU/KVM).

\subsection{Achieving Safety Objectives}
\label{sec:data-security}

\paragraph*{Host execution integrity}
By assuming that both the hypervisor and client-defined hypercall handlers (of
which there are few) are carefully implemented, using best practices of
software development, an adversarial virtine {\em cannot} directly modify the
state or code paths of the host. However, virtines {\em do not} guarantee that
if permitted access to certain hypercalls or secret data, an attacker cannot
utilize these hypercalls to exfiltrate sensitive data via side-channel
mechanisms. This, however, can be mitigated by using a mechanism that disables
certain hypercalls dynamically when they are not needed by the runtime, further
restricting the attack surface.

\paragraph*{Virtine execution integrity}
Requiring that no two virtines directly share memory without first receiving
permission from the hypervisor (e.g., via the hypercall interface) ensures data
secrecy within the virtine. Each virtine must have its own set of private data
which must be disjoint from any other virtine's set. Thus, a virtine that runs
untrusted or malicious code cannot affect the integrity of other virtines. 

\paragraph*{Isolation}
Modeling virtine and host private state as a disjoint set disallows any and all
shared state between virtines or the host. The hardware's use of nested paging
(EPT in VT-x) prevents such access at a hardware level. Also, by assuming that
hypercalls are carefully implemented, and that they only permit operations
required by the application, we achieve isolation from states and services
outside the virtine.

\section{Minimizing Virtine Costs}
\label{sec:costs}

Before exploring the implementation of virtines, we first describe a series of
experiments that guided their design. These experiments establish the creation
costs of minimal virtual contexts and of the execution environments used within
those contexts. Our goal is to establish what forms of overhead will be most
significant when creating a virtine.

\subsection{Experimental Setup} 
\label{sec:exp-setup}

The majority of our Linux and KVM experiments were run on {\em tinker}, an AMD
EPYC 7281 (Naples; 16 cores; 2.69\,GHz) machine with  32\,GB DDR4 running stock
Linux kernel version 5.9.12. We disabled hyperthreading, turbo boost, and DVFS
to mitigate measurement noise. We used a Dell XPS 9500 with an Intel i7 10750H
(Comet Lake; 6 cores) for SGX measurements. This machine has 32\,GB DDR4 and
runs stock Ubuntu 20.04 (kernel version 5.13.0-28). We used gcc 10.2.1 to
compile Wasp (C/C++), clang 10.0.1 for our C-based virtine language extensions,
and NASM v2.14 for assembly-only virtines. Unless otherwise noted, we conduct
experiments with 1000 trials. Note that our hypervisor implementation works on
both Linux and has a prototype implementation in Windows (through Hyper-V), but
for brevity we only show KVM's performance on Linux, as Hyper-V performance was
similar for our experiments.

\subsection{Measuring Startup Costs}
\label{sec:bootstrap-func}

\begin{figure}
    \centering
    \includegraphics[width=\columnwidth]{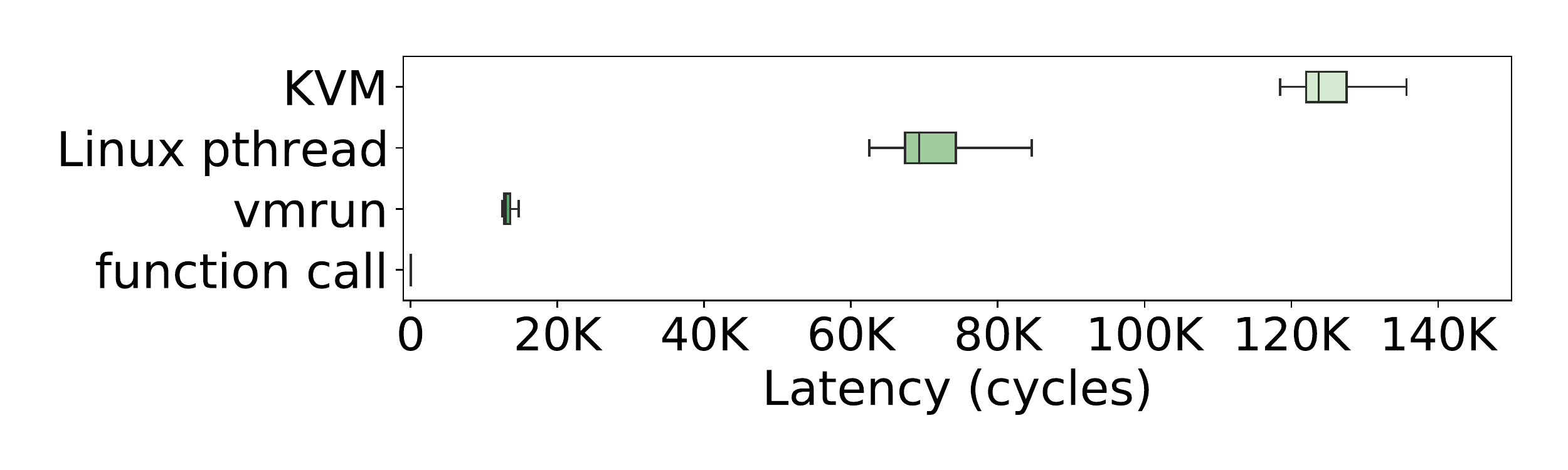}
    \caption{Lower bounds on execution context creation in cycles (measured with \texttt{rdtsc}).}
    \label{fig:abs-latency}
\end{figure}

We probe the costs of virtual execution contexts and see how they compare
to other types of execution contexts. To establish baseline creation costs, we
measure how quickly various execution contexts can be constructed on {\em
tinker}, as shown in Figure~\ref{fig:abs-latency}. We measure the time it takes
to create, enter, and exit from the context in a way that the hypervisor can
observe. In ``KVM'', we observe the latency to construct a virtual machine and
call the \verb|hlt| instruction. ``Linux pthread'' is simply a
\verb|pthread_create| call followed by \verb|pthread_join|. The ``vmrun''
measurement is the cost of running a VM hosted on KVM without the cost of
creating its associated state, i.e., only the \verb|KVM_RUN ioctl|. Finally,
``function'' is the cost of calling and returning from a null function. All
measurements are obtained using the \verb.rdtsc. instruction.

\paragraph*{Lower bounds}

The ``vmrun'' measurements represent the lowest latency we could
achieve to begin execution in a virtual context using KVM in Linux 5.9. This
latency includes the cost of the \verb.ioctl. system call, which in KVM is
handled with a series of sanity checks followed by execution of the
\verb.vmrun. instruction. Several optimizations can be made to the hypervisor
to reduce the cost of spawning new contexts and lower the latency of a virtine,
which we outline in Section~\ref{sec:cache-snapshot}.

These measurements tell us
that while a virtine invocation will be unsurprisingly more expensive than a
native function call, it can compete with thread creation and will far outstrip
any start-up performance that processes (and by proxy, containers) will achieve
in a standard Linux setting. We conclude that the baseline cost of {\em
creating} a virtual context is relatively inexpensive compared to the cost of
other abstractions.

\paragraph*{Eliminating traditional boot sequences}
The boot sequences of fully-featured OSes are too costly to include on the
critical path for low-latency function invocations~\cite{MANCO:2017:LIGHTVM,
CADDEN:2020:SEUSS, KOLLER:2017:SERVERLESS-UNIKERNELS}. It takes hundreds of
milliseconds to boot a standard Linux VM using QEMU/KVM. To understand why, we
measured the time taken for components of a vanilla Linux kernel boot sequence
and found that roughly 30\% of the boot process is spent scanning ACPI tables,
configuring ACPI, enumerating PCI devices, and populating the root file system.
Most of these features, such as a fully-featured PCI interface, or a network
stack, are unnecessary for short-lived, virtual execution environments, and are
often omitted from optimized Linux guest images such as the Alpine Linux image
used for Amazon's Firecracker~\cite{AGACHE:2020:FIRECRACKER-NSDI}.  Caching
pre-booted environments can further mitigate this overhead, as we describe in
\S\ref{sec:cache-snapshot}.

\begin{table}[]
\centering
    \begin{tabular}{@{}lrr@{}}
    \toprule
    \textbf{Component}            & \textbf{KVM} \\ \midrule
    Paging identity mapping       & 28109        \\ 
    Protected transition          & 3217         \\ 
    Long transition (\verb.lgdt.) & 681          \\ 
    Jump to 32-bit (\verb.ljmp.)  & 175          \\ 
    Jump to 64-bit (\verb.ljmp.)  & 190          \\ 
    Load 32-bit GDT (\verb.lgdt.) & 4118         \\ 
    First Instruction             & 74           \\ \bottomrule 
    \end{tabular}
\caption{Boot time breakdown for our minimal runtime environment on KVM. These
are minimum latencies observed per component, measured in cycles.} 
\label{tab:boot-breakdown}
\end{table}

In light of the data gathered in Figure~\ref{fig:abs-latency}, we set out to
measure the cost of creating a virtual context and configuring it with the
fewest operations possible. To do this, we built a simple wrapper
around the KVM interface that loads a binary image compiled from roughly 160
lines of assembly. This binary closely mirrors the boot sequence of a classic
OS kernel: it configures protected mode, a GDT, paging, and finally jumps to
64-bit code. These operations are outlined in Table~\ref{tab:boot-breakdown},
which indicates the minimum latencies (cycles) for each component, ordered by
cost.

The row labeled ``Paging/ident. map'' is by far the most expensive at $\sim$28K
cycles. Here we are using 2MB large pages to identity map the first 1GB of
address space, which entails three levels of page tables (i.e.,  12KB of memory
references), plus the actual installation of the page tables, control register
configuration, and construction of an EPT inside KVM. The transition to
protected mode takes the second longest, at 3K cycles. This is a bit
surprising, given that this only entails the protected mode bit flip (PE, bit
0) in \verb.cr0.. The transition to long mode (which takes several hundreds of
cycles) is less significant. The remaining components---loading a 32-bit GDT,
the long jumps to complete the mode transitions, and the initial interrupt
disable---are negligible.

\begin{figure}[]
    \centering
    \includegraphics[width=0.70\columnwidth]{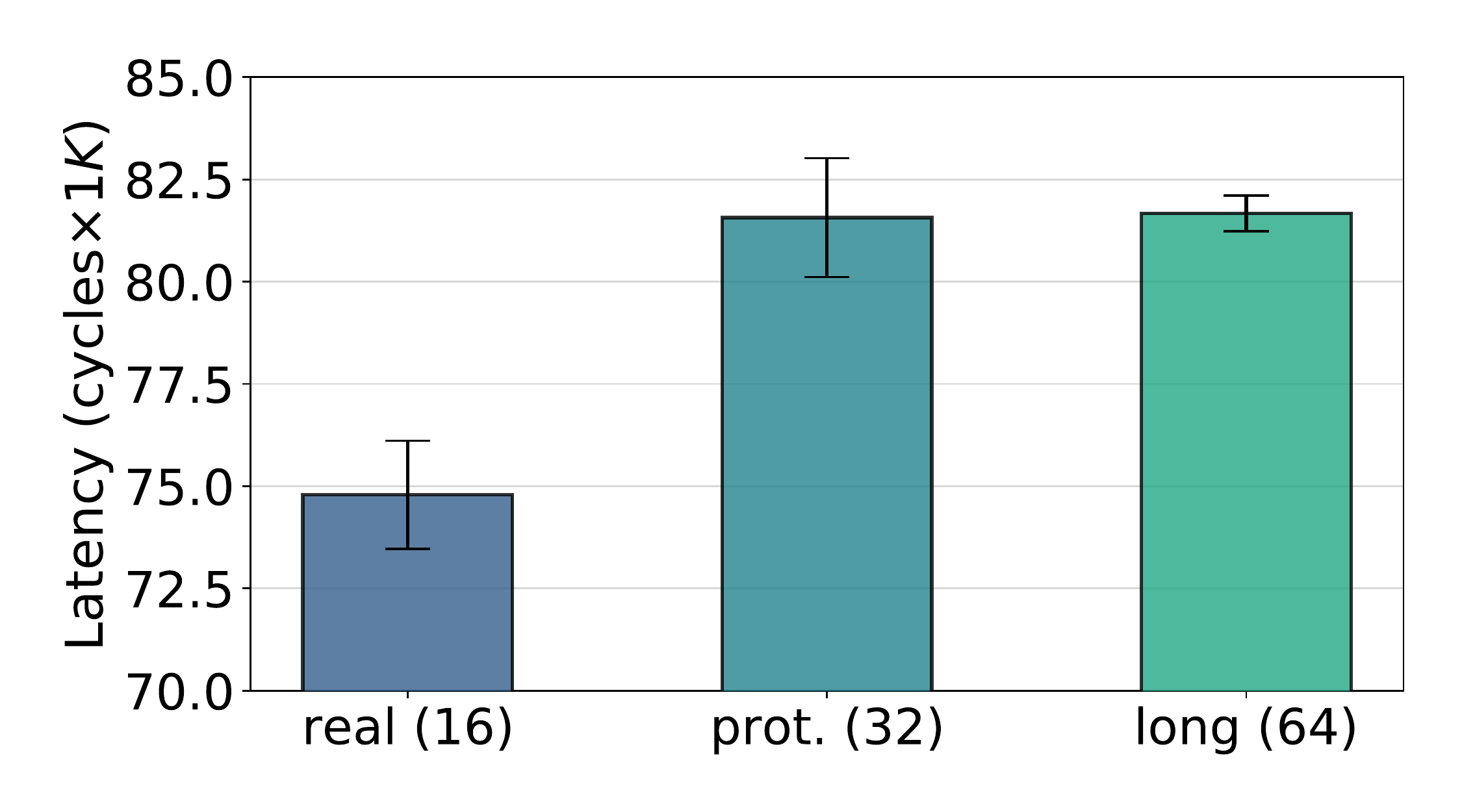}
    \caption{Latency to run a function in the three classic operating modes
    on x86. Note the use of a false origin to highlight relative differences.}
    \label{fig:x86}
\end{figure}

\paragraph*{The cost of processor modes}
The more complex the mode of execution (16, 32, or 64 bits), the higher the
latency to get there. This is consistent with descriptions in the hardware
manuals~\cite{INTEL_SYS_MAN, AMD_SYS_MAN}. To further investigate this effect,
we invoked a small binary written in assembly that brings the virtual context
up to a particular x86 execution mode and executes a simple function
(\verb|fib| of $20$ with a simple, recursive implementation).
Figure~\ref{fig:x86} shows our findings for the three canonical modes of the
x86 boot process using KVM: 16-bit (real) mode, 32-bit (protected) mode, and
64-bit (long) mode. Each mode includes the necessary components from
Table~\ref{tab:boot-breakdown} in the setup of the virtual context. In this
experiment, for each mode of execution, we measured the latency in cycles from
the time we initiated an entry on the host (\verb.KVM_RUN.), to the time it
took to bring the machine up to that mode in the guest (including the necessary
components listed in Table~\ref{tab:boot-breakdown}), run $fib(20)$, and exit
back to the host. These measurements include entry, startup cost, computation,
and exit. Note that we saw several outliers in all cases, likely due to host
kernel scheduling events. To make the data more interpretable, we removed these
outliers.\footnote{That is, using Tukey's method, measurements not on the
interval $[x_{25\%} - 1.5\,IQR, x_{75\%} + 1.5\,IQR]$ are removed from the
data.}

While we expect much of the time to be dominated by entry/exit and the
arithmetic, the benefits of real-mode only execution for our hand-written
version are clear. The difference between 16-bit and 32-bit environments are
not surprising. The most significant costs listed in
Table~\ref{tab:boot-breakdown} are not incurred when executing in 16-bit mode.
Protected and Long mode execution are essentially the same as they both include
those costs (paging and protected setup). These results suggest---provided that
the virtine is short-lived (on the order of \textit{microseconds}) and can
feasibly execute in real-mode---that 10K cycles may potentially be saved.



\begin{figure}[]
    \centering
    \includegraphics[width=\columnwidth]{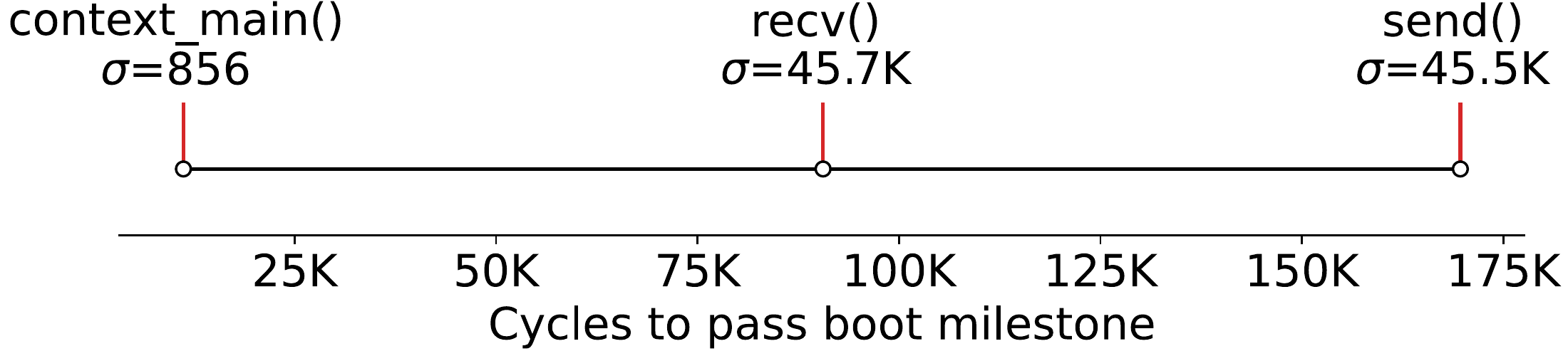}
		\caption{Latency for echo server startup milestones in protected mode (no paging).}
    \label{fig:echo-latency}
\end{figure}

\paragraph*{Booting up for useful code}
We have seen that a minimal long-mode boot sequence costs less than 30K cycles
($\sim$12 $\mu$s), but what does it take to do something useful? To determine
this, we implemented a simple HTTP echo server where each request is handled in
a new virtual context employing our minimal environment. We built a simple
micro-hypervisor in C++ and a runtime environment that brings the machine up to
C code and uses hypercall-based I/O to echo HTTP requests back to the sender.
The runtime environment comprises 970 lines of C (a large portion of which are
string formatting routines) and 150 lines of x86 assembly.  The
micro-hypervisor comprises 900 lines of C++. The hypercall-based I/O (described
more in Section~\ref{sec:wasp}) obviates the need to emulate network devices in
the micro-hypervisor and implement the associated drivers in the virtual
runtime environment, simplifying the development process.
Figure~\ref{fig:echo-latency} shows the mean time measured in cycles to pass
important startup milestones during the bring-up of the server context.  The
left-most point indicates the time taken to reach the server context's main
entry point (C code); roughly 10K cycles.  Note that this example does not
actually require 64-bit mode, so we omit paging and leave the context in
protected mode. The middle point shows the time to receive a request (the
return from \verb.recv().), and the last point shows the time to complete the
response (\verb.send().). Milestone measurements are taken inside the virtual
context.

The send and receive functions for this environment use hypercalls to defer to
the hypervisor, which proxies them to the Linux host kernel using the
appropriate system calls. Even when leveraging the underlying host OS, and when
adding the from-scratch virtual context creation time from
Figure~\ref{fig:abs-latency}, we can achieve sub-millisecond HTTP response
latencies ($<$300 $\mu$s) {\em without optimizations}
(\S\ref{sec:cache-snapshot}). Thus, we can infer that despite the cost of
creating a virtual context, having few host/virtine interactions can keep
execution latencies in a virtual context within an acceptable range. Note,
however, that the guest-to-host interactions in this test introduce variance
from the host kernel's network stack, indicated by the large standard
deviations shown in Figure~\ref{fig:echo-latency}.

These results are promising, and they indicate that we can achieve low
overheads and start-up latencies for functions that do not require a
heavy-weight runtime environment. We use three key insights from this section
to inform the design of our virtine framework in the next section: (1) creating
hardware virtualized contexts can be cheap when the environment is small, (2)
tailoring the execution environment (for example, the processor mode) can pay
off, and (3) host interactions can be facilitated with hypercalls (rather than
shared memory), but their number must be limited to keep costs low.

\section{Implementation}
\label{sec:impl}

In this section, we present Wasp, a prototype hypervisor designed for the
creation and management of virtine environments. We also cover a few of the
optimizations designed to overcome the cost of creating virtual contexts using
KVM.

\subsection{Wasp}
\label{sec:wasp}

\begin{figure}
    \centering
    \includegraphics[width=0.9\columnwidth]{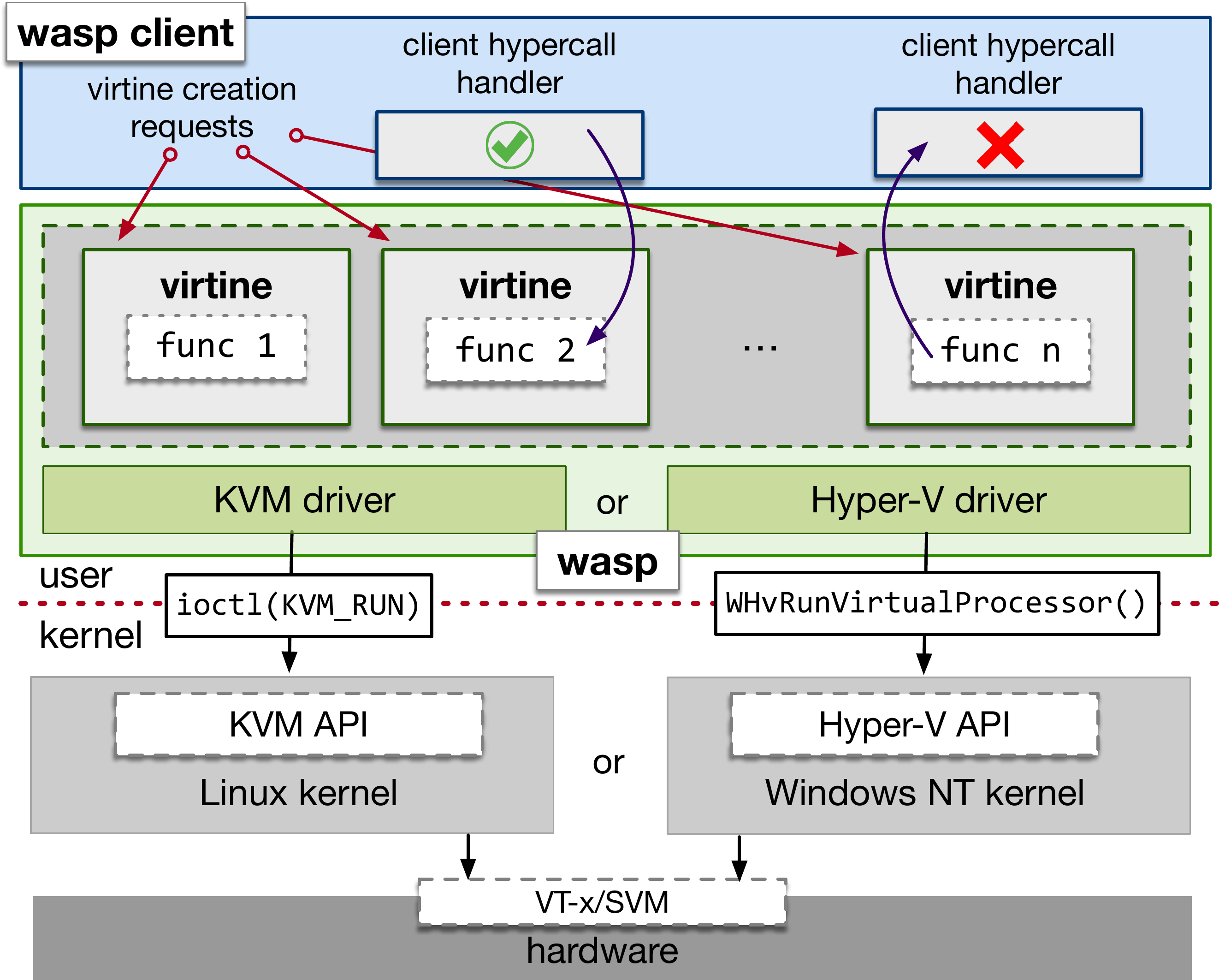}
    \caption{High-level overview of Wasp.}
    \label{fig:arch-diag}
\end{figure}

Wasp is a specialized, embeddable micro-hypervisor runtime that deploys
virtines with an easy-to-use interface.  Wasp runs on Linux and Windows. At its
core, Wasp is a fairly ordinary hypervisor, hosting many virtual contexts on
top of a host OS.  However, like other minimal hypervisors such as
Firecracker~\cite{AGACHE:2020:FIRECRACKER-NSDI}, Unikernel
monitors~\cite{WILLIAMS:2016:UKM}, and uhyve~\cite{LANKES:2017:UHYVE}, Wasp
does not aim to emulate the entire x86 platform or device model. As shown in
Figure~\ref{fig:arch-diag}, Wasp is a userspace runtime system built as a
library that host programs (virtine clients) can link against. Wasp mediates
virtine interactions with the host via a hypercall interface, which is checked
by the hypervisor and the virtine client. The figure shows one virtine that has
no host interactions, one virtine which makes a valid hypercall request, and
another whose hypercall request is denied by the client-specified security
policy. By using Wasp's runtime API, a virtine client can leverage hardware
specific virtualization features without knowing their details. Several types of applications
(including dynamic compilers and other runtime systems) can
link with the Wasp runtime library to leverage virtines. On Linux, each virtual
context is represented by a device file which is manipulated by Wasp using an
\verb$ioctl$.

Wasp provides no libraries to the binary being run, meaning they have no
in-virtine runtime support by default.  Wasp simply accepts a binary image,
loads it at guest virtual address \verb.0x8000., and enters the VM context. Any
extra functionality must be achieved by interacting with the hypervisor and
virtine client. In Wasp, delegation to the client is achieved with hypercalls
using virtual I/O ports.

Hypercalls in Wasp are not meant to emulate low-level virtual devices, but are
instead designed to provide high-level hypervisor services with as few exits as
possible. For example, rather than performing file I/O by ultimately
interacting with a virtio device~\cite{RUSSEL:2008:VIRTIO} and parsing
filesystem structures, a virtine could use a hypercall that mirrors the
\verb|read| POSIX system call. Hypercalls vector to a co-designed handler either
provided by Wasp or implemented by the virtine client. Wasp provides the mechanisms to create
virtines, while the client can specify security policies through handlers.
These handlers could simply run a series of checks and pass through certain
host system calls while filtering others out. While virtine clients
can implement custom hypercall handlers, they can also choose from a variety of
general-purpose handlers that Wasp provides out-of-the-box; these canned
hypercalls are used by our language extensions (\S\ref{sec:clang}). 
By default, Wasp provides no externally observable behavior through hypercalls other than
the ability to exit the virtual context; all other external behavior must be
validated and expressly permitted by the custom (or canned) hypercall handlers, which are implemented
(or selected) by the virtine client. 


\begin{figure}
    \centering
    \includegraphics[width=0.9\columnwidth]{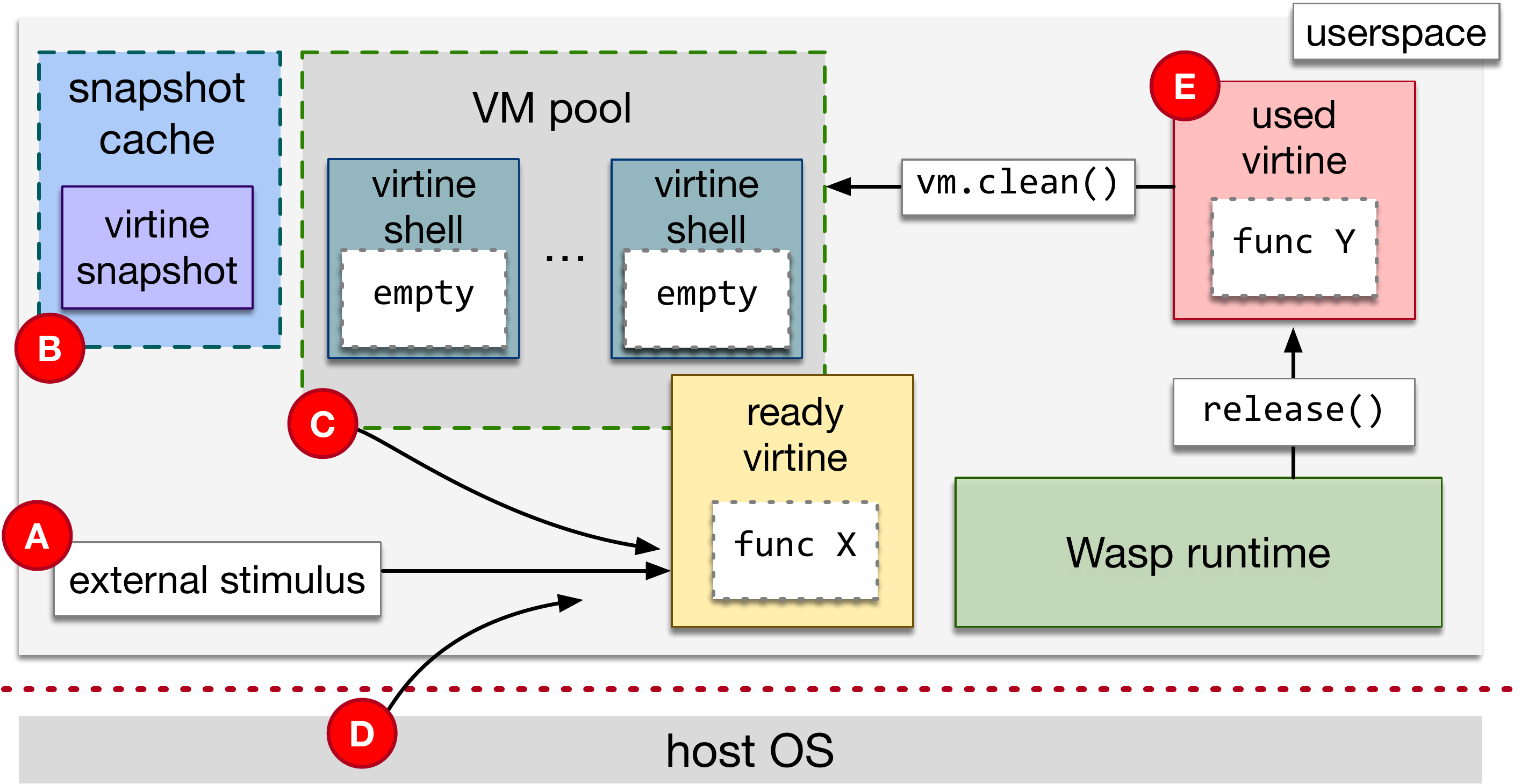}
    \caption{Image snapshotting and virtine reuse with a pooled design.}
    \label{fig:pool-design}
\end{figure}

\subsection{Wasp Caching and Snapshotting}
\label{sec:cache-snapshot}

\paragraph*{Caching}
To reduce virtine start-up latencies, Wasp supports a pool of cached,
uninitialized, virtines (shells) that can be reused. As depicted in
Figure~\ref{fig:pool-design}, Wasp receives a request from a virtine client
(\rcwhite{A}), which will drive virtine creation. Such requests can be generated in a variety
of virtine client scenarios. For example, network traffic hitting a web server that
implements a virtine client may generate virtine invocations. A database engine incorporating a virtine client may run virtine-based
UDFs in response to triggers. Because we must use a new virtine for
every request, a hardware virtual context must be provisioned to handle each
invocation. The context is acquired by one of two methods, provisioning a clean
virtual context (\rcwhite{C}) or reusing a previously created context (\rcwhite{D}). 
When the system is cold (no virtines have yet been created), we
must ask the host kernel for a new virtual context by using KVM's
\verb.KVM_CREATE_VM. interface.  If this route is taken, we pay a higher cost
to construct a virtine due to the host kernel's internal allocation of the VM
state (VMCS on Intel/VMCB on AMD). However, once we do this, and the relevant
virtine returns, we can clear its context (\rcwhite{E}), preventing information
leakage, and cache it in a pool of ``clean'' virtines (\rcwhite{C}) so the host
OS need not pay the expensive cost of re-allocating virtual hardware contexts.
These virtine ``shells'' sit dormant waiting for new virtine creation requests
(\rcwhite{B}). The benefits of pooling virtines are apparent in
Figure~\ref{fig:wasp-latency} by comparing creation of a Wasp virtine from
scratch (the ``Wasp'' measurement) with reuse of a cached virtine shell from
the pool (``Wasp+C''). By recycling virtines, we can reach latencies much lower
than Linux thread creation and much closer to the hardware limit, i.e., the
\verb.vmrun. instruction. Note that here we include Linux process creation
latencies as well for scale. Included is the ``Wasp+CA'' (cached, asynchronous)
measurement, which does not measure the cost of cleaning virtines and instead
cleans them asynchronously in the background. This can be implemented by either
a background thread or can be done when there are no incoming requests. This
measurement shows that the caching mechanism brings the cost of
provisioning a virtine shell to within $4\%$ of a bare \verb.vmrun..

We also measured these costs on a recent SGX-enabled Intel platform and
observed similar behavior, as shown in the bottom half of
Figure~\ref{fig:wasp-latency}. The ``SGX Create'' measurement indicates the
cost of creating a new enclave, and the ECALL measurement indicates the cost of
\textit{entering} an enclave, thus reusing the previously created context.

\begin{figure}
    \centering
    \includegraphics[width=0.8\columnwidth]{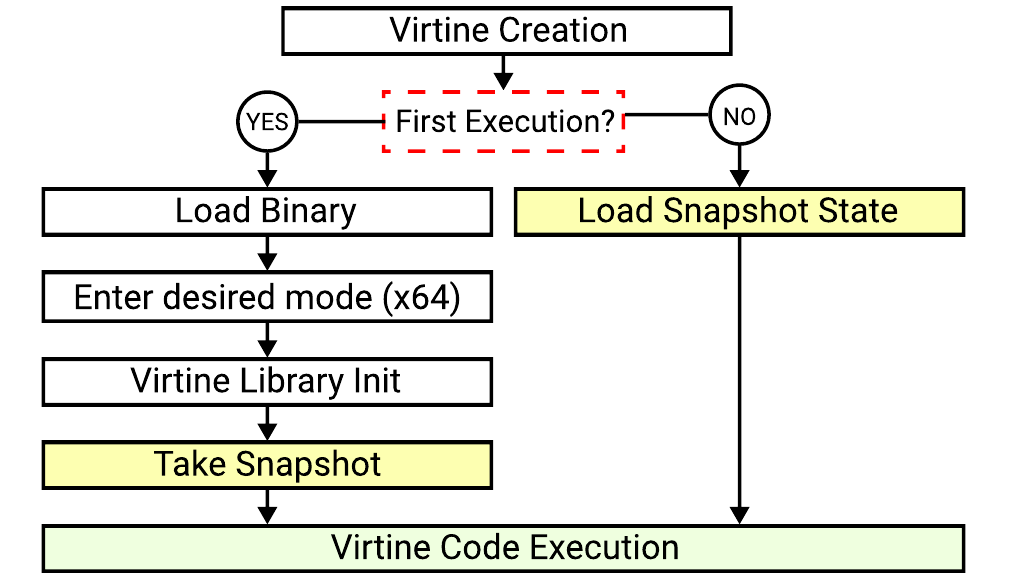}
    \caption{Virtine code path on first execution and subsequent executions after snapshotting.}
    \label{fig:snapshot-timeline}
\end{figure}

\paragraph*{Snapshotting}

As was shown in Section~\ref{sec:costs}, the initialization of a virtine's
execution state can lead to significant overheads compared to traditional
function calls. This overhead is undesirable if the code that is executed in a
virtine is not particularly long-lived (less than a few microseconds). Others
have mitigated these start-up latencies in the serverless domain by
``checkpointing'' or ``snapshotting'' container runtime state after
initialization~\cite{OAKES:2018:SOCK,DU:2020:CATALYZER, CADDEN:2020:SEUSS}. In
a similar fashion, Wasp supports snapshotting by allowing a virtine to leverage
the work done by previous executions of the same function. As outlined in
Figure~\ref{fig:snapshot-timeline}, the first execution of a virtine must still
go through the initialization process by entering the desired mode and
initializing any runtime libraries (in this case, libc). The virtine then takes
a snapshot of its state, and continues executing. Subsequent executions of the
same virtine can then begin execution at the snapshot point and skip the
initialization process. This optimization significantly reduces virtine
overheads, which we explore further in Section~\ref{sec:clang}. Of course, by
snapshotting a virtine's private state, that state is exposed to all future
virtines that are created using that ``reset state.'' Thus, care must be taken
in describing what memory is saved in a snapshot in order to maintain the
isolation objectives outlined in Section~\ref{sec:data-security}. We detail the
costs involved in snapshotting in Section~\ref{sec:img-size}.

\begin{figure}
    \raggedright
    \includegraphics[width=\columnwidth]{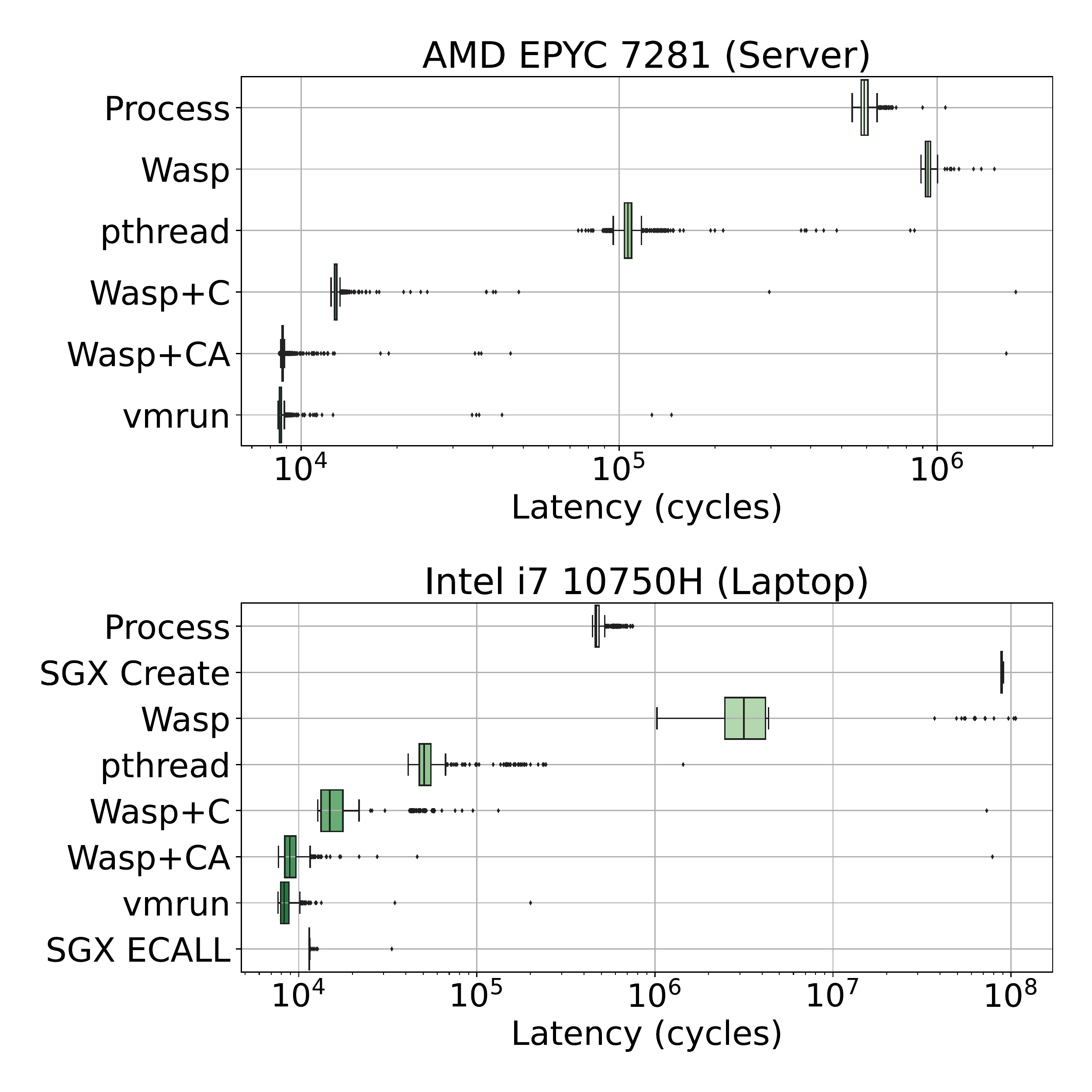}
		\caption{Creation latencies for execution contexts on modern AMD and Intel platforms, including Wasp virtines and SGX where available. Note the log scale on the horizontal axis.}
    \label{fig:wasp-latency}
\end{figure}

\subsection{C Language Extensions}
\label{sec:clang}

While Wasp significantly eases the development and deployment of virtines, with
only the runtime library, developers must still manage virtine internals,
namely the build process for the virtine's internal execution environment.
Requiring developers to create kernel-style build systems that package boot
code, address space configurations, a minimal libc, and a linker script per
virtine creates an undue burden. To alleviate this burden, we implemented a
clang wrapper and LLVM compiler pass. The purpose of the clang wrapper is to 
include our pass in the invocation of the middle-end. The compiler pass detects C functions annotated with
the \texttt{virtine} keyword, runs middle-end analysis at the IR level, and
automatically generates code that invokes a pre-compiled virtine binary
whenever the function is called. When this pass detects a function annotation
as shown in Figure~\ref{fig:c-virtine-api}, it generates a call graph rooted at
that function. The compiler automatically packages a subset of the source
program into the virtine context based on what that virtine needs. Global
variables accessed by the virtine are currently initialized with a snapshot
when the virtine is invoked. Concurrent modifications (e.g., by different
virtines, or by the client and a virtine) will occur on distinct copies of the
variable. Currently, if a virtine calls another virtine-annotated function, a nested
virtine will not be created. 

\begin{figure}[]\footnotesize
\begin{tabular}{c}
\begin{lstlisting}[
    language=C,
    showstringspaces=false]
virtine int fib(int n) {
    if (n < 2) return n;
    return fib(n - 1) + fib(n - 2);
}
\end{lstlisting}
\end{tabular}
    \caption{Virtine programming in C with compiler support.}
\label{fig:c-virtine-api}
\end{figure}

To further ease programming burden, compiler-supported virtines must have
access to some subset of the C standard library. Due to the nature of their
runtime environment, basic virtines do not include these libraries.  To
remedy this, we created a virtine-specific port of {\em newlib}~\cite{NEWLIB},
an embeddable C standard library that statically links and maintains a
relatively small virtine image size. Newlib allows developers to provide their
own system call implementations; we simply forward them to the hypervisor as a
hypercall.  When the \verb|virtine| keyword is used, all hypercalls are restricted
by default,
following the default-deny semantics of virtines previously mentioned. If,
however, the programmer (implementing the virtine client) would like to permit
hypercalls, they can use the \verb|virtine_permissive| keyword to allow all
hypercalls, or the  \verb|virtine_config(cfg)| to supply a configuration
structure that contains a bit mask of allowed hypercalls. If a hypercall is
permitted, the handler in the client must validate the arguments and service
it, for example by delegating to the host kernel's system call interface or by
performing client-specific emulation.

This allows virtines to support standard library functionality without
drastically expanding the virtine runtime environments. Of course, by using a
fully fledged standard library, the user still opens themselves up to common
programming errors. For example, an errant \verb|strcpy| can still result in
undefined (or malicious) behavior, but this has no consequences for the host or
other virtines as outlined in Section~\ref{sec:data-security}. All virtines
created via our language extensions use Wasp's snapshot feature by default.
This can be disabled with the use of an environment variable.

\subsection{Execution Environments}
\label{sec:exec-environ}

\begin{figure}
    \centering
    \includegraphics[width=0.9\columnwidth]{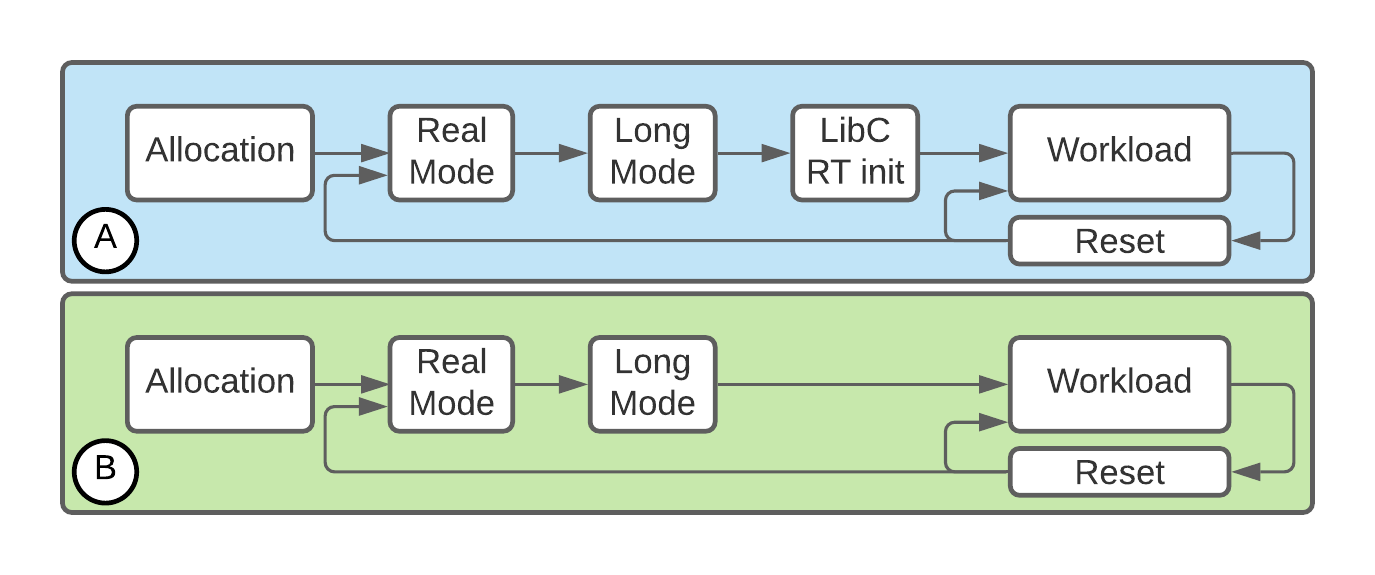}
    \caption{Default execution environments available to a virtine.}
    \label{fig:exec-environs}
\end{figure}

Wasp provides two default execution environments for programmers to use, though
others are possible. These default environments are shown in
Figure~\ref{fig:exec-environs}. For the C extensions (\wcblack{A}), the virtine
is pre-packaged with a POSIX-like runtime environment, which stands between the
``boot'' process and the virtine's function. If a programmer directly uses the
Wasp C++ API, (\wcblack{B}), the virtine is not automatically packaged with a
runtime, and it is up to the client to provide the virtine binary. Both
environments can use snapshotting after the reset stage, allowing them to skip
the costly boot sequence. We envision an environment management system that
will allow programmers to treat these environments much like package
dependencies~\cite{LIU:2019:DIVER}.

\section{Evaluation}
\label{sec:eval}

In this section, we evaluate virtines and the Wasp runtime using microbenchmarks and case studies
that are representative of function isolation in the wild. 
With these experiments, we seek to answer the following questions:

\BEGINITEMIZE
\item How significant are baseline virtine startup overheads with our language extensions, and how much computation is
    necessary to amortize them? (\S\ref{sec:fib})
\item What is the impact of the virtine's execution environment (image size) on start-up cost? (\S\ref{sec:img-size})
\item What is the performance penalty for host interactions? (\S\ref{sec:http})
\item How much effort is required to integrate virtines with off-the-shelf library code? (\S\ref{sec:openssl})
\item How difficult is it to apply virtines to managed language use cases
    and what are the costs? (\S\ref{sec:js})
\ENDITEMIZE

\begin{figure}[]
    \centering
    \includegraphics[width=0.9\columnwidth]{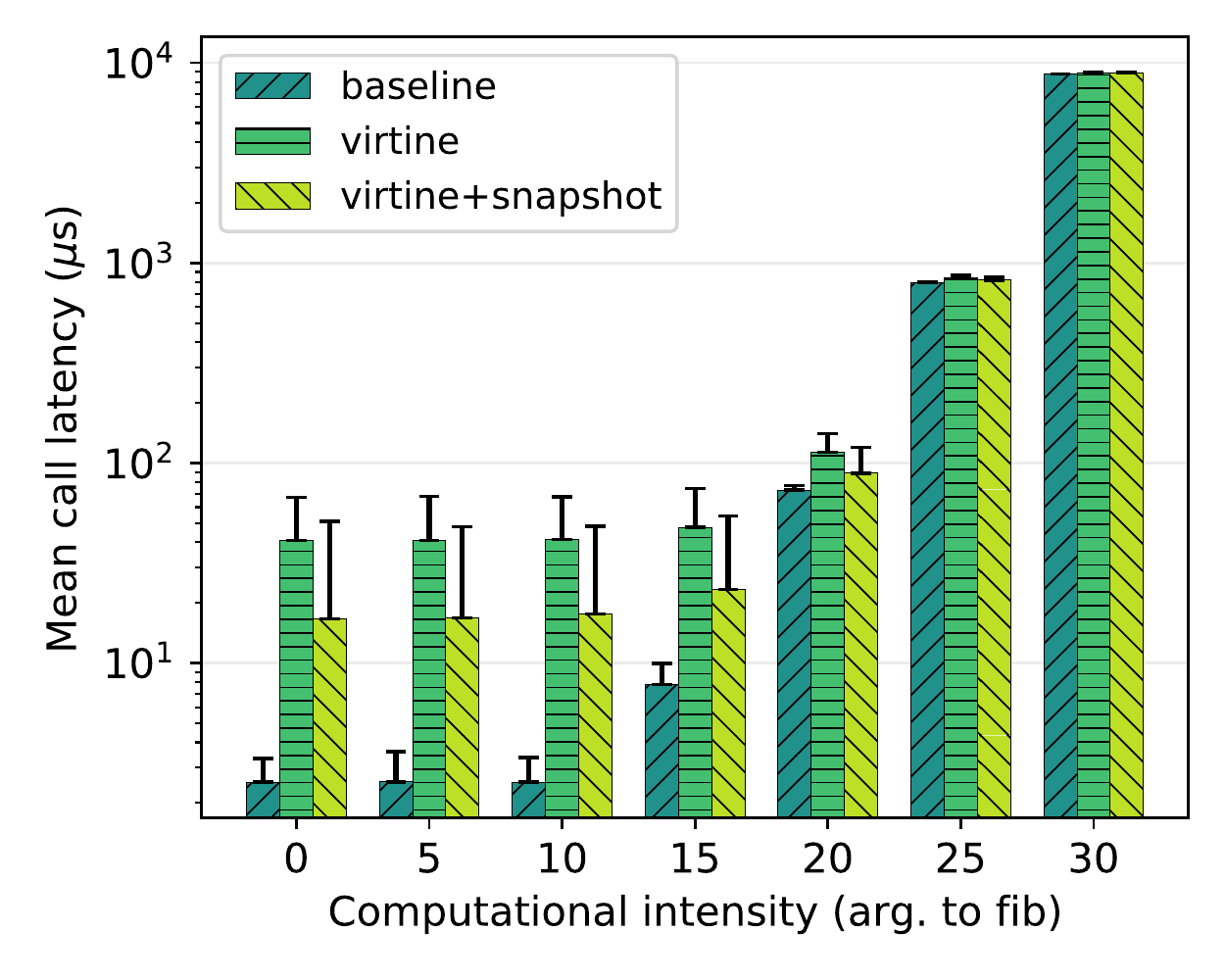}
    \caption{Latency of virtines as computational intensity increases. Note the
             log scale on the vertical axis.}
    \label{fig:fib-latency}
\end{figure}

\subsection{Startup Latencies with Language Extensions}
\label{sec:fib}

We first study the start-up overheads of virtines using our language
extensions. We implemented the minimal \verb|fib| example shown in
Figure~\ref{fig:c-virtine-api} and scaled the argument to \verb|fib| to
increase the amount of computation per function invocation, shown in
Figure~\ref{fig:fib-latency}. We compare virtines with and without image
snapshotting to native function invocations. $fib(0)$ essentially measures the
inherent overhead of virtine creation, and as $n$ increases, the cost of
creating the virtine is amortized. The measurements include setup of a basic
virtine image (which includes libc), argument marshalling, and minimal machine
state initialization. The argument, $n$, is loaded into the virtine's address
space at address \verb.0x0.. In the case of the experiment labeled ``virtine +
snapshot,'' a snapshot of the virtine's execution state is taken on the first
invocation of the \verb|fib| function. All subsequent invocations of that
function will use this snapshot, skipping the slow path boot sequence (see
Figure~\ref{fig:snapshot-timeline}) producing an overall speedup of $2.5\times$
relative to virtines \textit{without} snapshotting for $fib(0)$. Note that we
are not measuring the steady state, so the bars include the overhead for taking
the initial snapshot. This is why we see more variance for the snapshotting
measurements. 

At first, the relative slowdown between native function invocation and virtines
with snapshotting is $6.6\times$. When the virtine is short-lived, the costs of
provisioning a virtine shell and initializing it account for most of the
execution time. However, with larger computational requirements, the slowdown
drops to $1.03\times$ for $n=25$ and $1.01\times$ for $n=30$. This shows that
as the function complexity increases, virtine start-up overheads become
negligible, as expected. Here we can amortize start-up overheads with
$\sim$100$\mu$s of work. 

\begin{table}[]
    \centering
        \begin{tabular}{@{}lll@{}}
        \toprule
        \textbf{System} & \textbf{Latency}   &   \textbf{Boundary Cross Mechanism}  \\ \midrule
            Wedge~\cite{BITTAU:2008:WEDGE}             &   $\sim$60$\mu$s     &   {\em sthread} call \\ 
            LwC~\cite{LITTON:2016:LWC}             &  2.01$\mu$s        &   \verb.lwSwitch.  \\ 
            Enclosures~\cite{GHOSN:2021:ENCLOSURES}      &   0.9$\mu$s        &   Custom syscall interface  \\ 
            SeCage~\cite{LIU:2015:SECAGE}          &   0.5$\mu$s        &   VMRUN/VMFUNC   \\ 
            Hodor~\cite{HEDAYATI:2019:HODOR}           &   0.1$\mu$s        &   VMRUN/VMFUNC  \\
        Virtines        &   5$\mu$s          &   Syscall interface + VMRUN  \\ \bottomrule
    \end{tabular}
    \caption{Comparing costs of crossing isolation boundaries.}
    \label{tab:entering-isolations}
\end{table}

We compare virtine start-up costs to the cost of crossing isolation boundaries
in other published systems in Table~\ref{tab:entering-isolations}. While the
types of isolation these systems provide is slightly different, these numbers
put the cost of the underlying mechanism into perspective. LwC and Enclosures
switch between isolated contexts within the same kernel in a similar way to
process-based isolation. SeCage and Hodor measure only the latency of the
VMFUNC instruction without a VMEXIT event. Virtine latency is measured from
userspace on the host, surrounding the KVM\_RUN ioctl, thus incurring
system call and ring-switch overheads.

\subsection{Impact of Image Size}
\label{sec:img-size}

\begin{figure}
    \centering
    \includegraphics[width=\columnwidth]{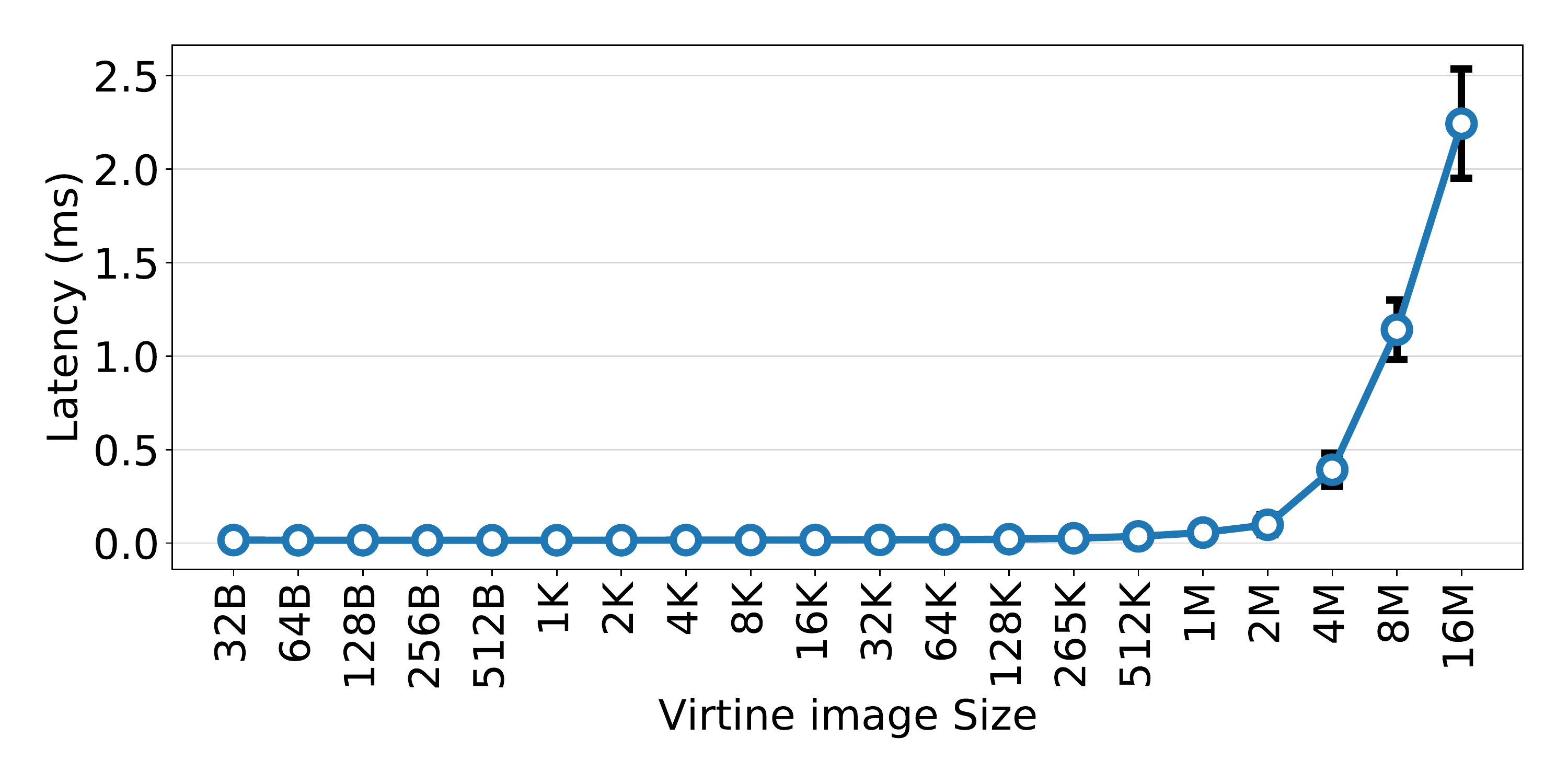}
    \caption{Impact of image size on start-up latency.}
    \label{fig:image-size}
\end{figure}

To evaluate the impact of virtines' execution environments on start-up costs,
we performed an experiment that artificially increases image size, shown in
Figure~\ref{fig:image-size}. This figure shows increasing virtine image size
(up to 16MB) versus virtine execution latency for a minimal virtine that simply
halts on startup. We synthetically increase image size by padding
a minimal virtine image with zeroes. With a 16MB image size, the start-up cost
is 2.3ms. This amounts to roughly 6.8GB/s, which is in line with our
measurement of the \verb|memcpy| bandwidth on our \textit{tinker} machine, 6.7GB/s.  This
shows the minimal cost a virtine will incur for start-up with a simple
snapshotting strategy when the boot sequence is eliminated. Using
a copy-on-write approach, as is done in SEUSS~\cite{CADDEN:2020:SEUSS}, we
expect this cost could be reduced drastically.

These results reflect what others have seen for unikernel boot times.
Unikernels tend to have a larger image size than what would be needed for a
virtine execution environment, and thus incur longer start-up times. Kuenzer et
al. report the shortest we have seen, at 10s to 100s of $\mu$s for
Unikraft~\cite{KUENZER:2021:UNIKRAFT}, while other unikernels
(MirageOS~\cite{MADHAVAPEDDY:2013:UNIKERNELS}, OSv~\cite{KIVITY:2014:OSV},
Rump~\cite{KANTEE:2012:RUMP}, HermiTux~\cite{OLIVIER:2019:HERMITTUX}, and
Lupine~\cite{KUO:2020:LUPINE}) take tens to hundreds of milliseconds to boot a
trivial image. For example, we measured the no-op function evaluation time
under OSv to be roughly 600 milliseconds on our testbed. A similar no-op
function achieved roughly 12ms under MirageOS run with Solo5's HVT
tender~\cite{SOLO5}, which directly
interfaces with KVM and uses hypercalls in a similar way to virtines.

\subsection{Host Interaction Costs}
\label{sec:http}

As outlined in Section~\ref{sec:virtines}, virtines must interact with the
client for all actions that are not fulfilled by the environment within the
virtine. For example, a virtine must use hypercalls to read files or access
shared state. Here we attempt to determine how frequent client interactions
(via hypercalls) affect performance for an easily understood example. To do so,
we use our C extension to annotate a connection handling function in a simple,
single-threaded HTTP server that serves static content. Each connection that
the server receives is passed to this function, which automatically provisions
a virtine environment. 

\begin{figure}
    \begin{centering}
    \subfigure[]{
        \includegraphics[width=0.47\columnwidth]{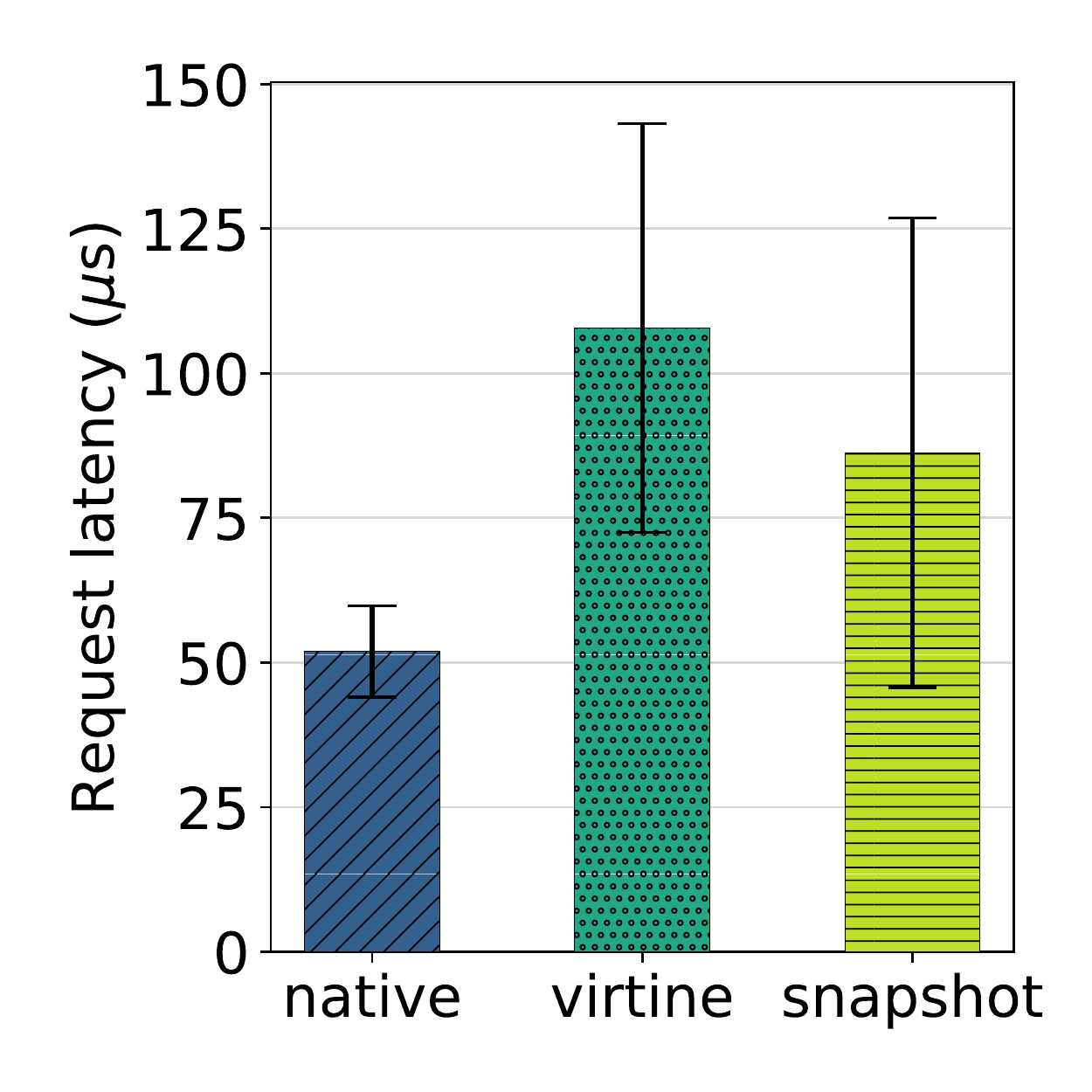}
        \label{fig:http-latency}
    }
    \subfigure[]{
        \includegraphics[width=0.47\columnwidth]{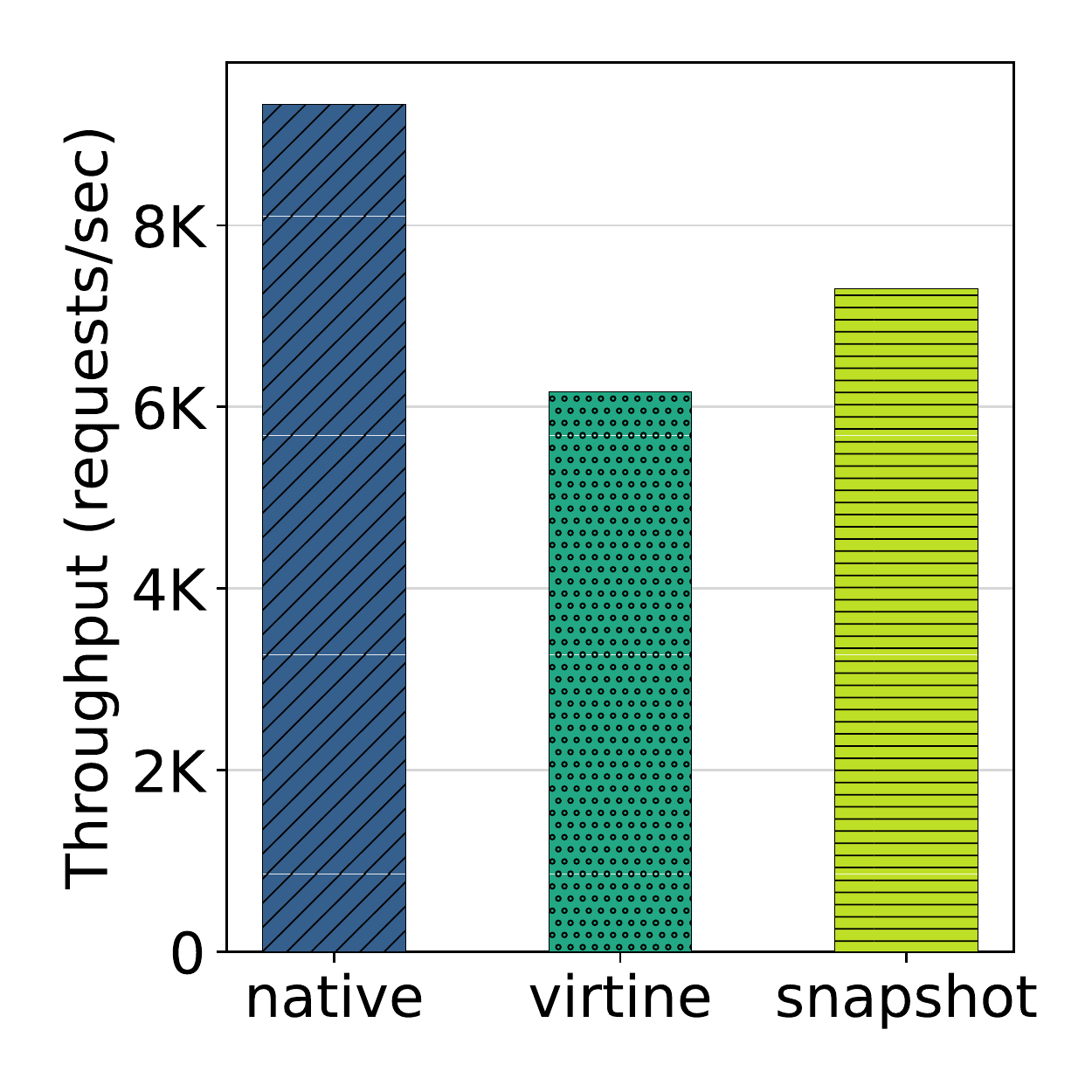}
        \label{fig:http-tput}
    }
    \end{centering}
    \caption{Mean response latency (a) and harmonic mean of throughput (b) for a simple HTTP server written in C, with each
    request handled natively and in a virtine (with and without snapshotting).}
    \label{fig:http-perf}
\end{figure}

We measured both the latency and throughput of HTTP requests with and without
virtines on {\em tinker}. The results are shown in Figure~\ref{fig:http-perf}.
Virtine performance is shown with and without snapshotting (``virtine'' and
``snapshot''). Requests are generated from \verb|localhost| using a custom
request generator (which always requests a single static file).  Note that each
virtine invocation here involves seven host interactions (hypercalls): (1)
\verb|read()| a request from host socket, (2) \verb|stat()| requested file, (3)
\verb|open()| file, (4) \verb|read()| from file, (5) \verb|write()| response,
(6) \verb|close()| file, (7) \verb|exit()|. Wasp handles these hypercalls by
first validating arguments, and if they are allowed through, re-creates the
calls on the host. For example, a validated \verb|read()| will turn into a
\verb|read()| on the host filesystem. The exits generated by these hypercalls
are doubly expensive due to the ring transitions necessitated by KVM. However,
despite the cost of these host interactions, virtines with snapshots incur only
a 12\% decrease in throughput relative to the baseline. We expect that these
costs would be reduced in a more realistic HTTP server, as more work unrelated
to I/O would be involved. This effect has been observed by others employing
connection sandboxing~\cite{GHOSN:2021:ENCLOSURES}.

\subsection{Integration with Library Code}
\label{sec:openssl}

To investigate the difficulty of incorporating virtines into libraries, and
more significant codebases, we modified off-the-shelf
OpenSSL.\footnote{OpenSSL version 3.0.0 alpha7.} OpenSSL is used as a library
in many applications, such as the Apache web server, Lighttpd, and OpenVPN.
We changed the library so that its 128-bit AES block cipher encryption is
carried out in virtine context. We chose this function since it is a core
component of many higher-level encryption features. While this would not be a
good candidate for running in virtine context from a performance perspective,
it gives us an idea of how difficult it is to use virtines to isolate a deeply
buried, heavily optimized function in a large codebase.

Compiling OpenSSL using virtines was straightforward.  From the
developer's perspective, it simply involved annotating the block cipher
function with the \verb.virtine. keyword and integrating our custom clang/LLVM
toolchain with the OpenSSL build environment (i.e., swapping the default
compiler).  The latter step was more work. In all, the change took roughly one
hour for an experienced developer. 


Though our main goal here was not to evaluate end-to-end
performance, we did measure the performance impact of integrating virtines
using OpenSSL's internal benchmarking tool.  We ran the built-in speed
benchmark\footnote{\texttt{openssl speed -elapsed -evp aes-128-cbc}} to measure
the throughput of the block cipher using virtines (with our snapshotting
optimization) compared to the baseline (native execution).  Note that since the
block cipher is being invoked many thousands of times per second, virtine
creation overheads amplify the invocation cost significantly. In a realistic
scenario, the developer would likely include more functionality in virtine
context, amortizing those overheads. That said, with our optimizations and a
16KB cipher block size, virtines only incur a 17$\times$ slowdown relative to
native execution with snapshotting. The OpenSSL virtine image we use is
roughly 21KB, which following Figure~\ref{fig:image-size} will translate to
16$\mu$s for every virtine invocation. It follows, then, that virtine creation
in this example is memory bound, since copying the snapshot comprises the
dominant cost.

\subsection{Virtines for Managed Languages}
\label{sec:js}

As described in our threat model (\S\ref{sec:threat}), virtines can provide
isolation in environments where untrusted code executes. Examples
of such environments are serverless platforms and databases UDFs. These
environments often use high-level languages like JavaScript, Python, or Java
to isolate the untrusted code. However, this isolation can
still be compromised by bugs in the isolation logic.

Motivated by these environments, we investigate how a managed language can incorporate
fine-grained isolation by running JavaScript functions in virtine context,
and by exploring how virtine-specific optimizations can be used to reduce costs
and improve latencies.

\paragraph*{Implementation}
We chose the Duktape JavaScript engine for
its portability, ease-of-use, and small memory footprint~\cite{DUKTAPE}. Our baseline
implementation (no virtines) is configured to allocate a Duktape context,
populate several native function bindings, execute a function that base64
encodes a buffer of data, and returns the encoding to the caller after tearing
down (freeing) the JS engine. The virtine does the same thing, but uses the Wasp
runtime library directly (no language extensions). This allows the engine to use
only three hypercalls: \verb|snapshot()|, \verb|get_data()|, and
\verb|return_data()|. The snapshot hypercall instructs the runtime to take a
snapshot after booting into long mode and allocating the Duktape context.
\verb.get_data(). asks the hypervisor to fill a buffer of memory with the data
to be encoded, and once the virtine encodes the data, it calls
\verb.return_data(). and the virtine exits. By co-designing the hypervisor and
the virtine, and by providing only a limited set of hypercalls, we limit the attack
surface available to an adversary. For example, \verb.snapshot. and
\verb.get_data. cannot be called more than once, meaning that if an attacker
were to gain remote code execution capabilities, the only permitted hypercall
would terminate the virtine.

\begin{figure}
\centering
    \includegraphics[width=0.9\columnwidth]{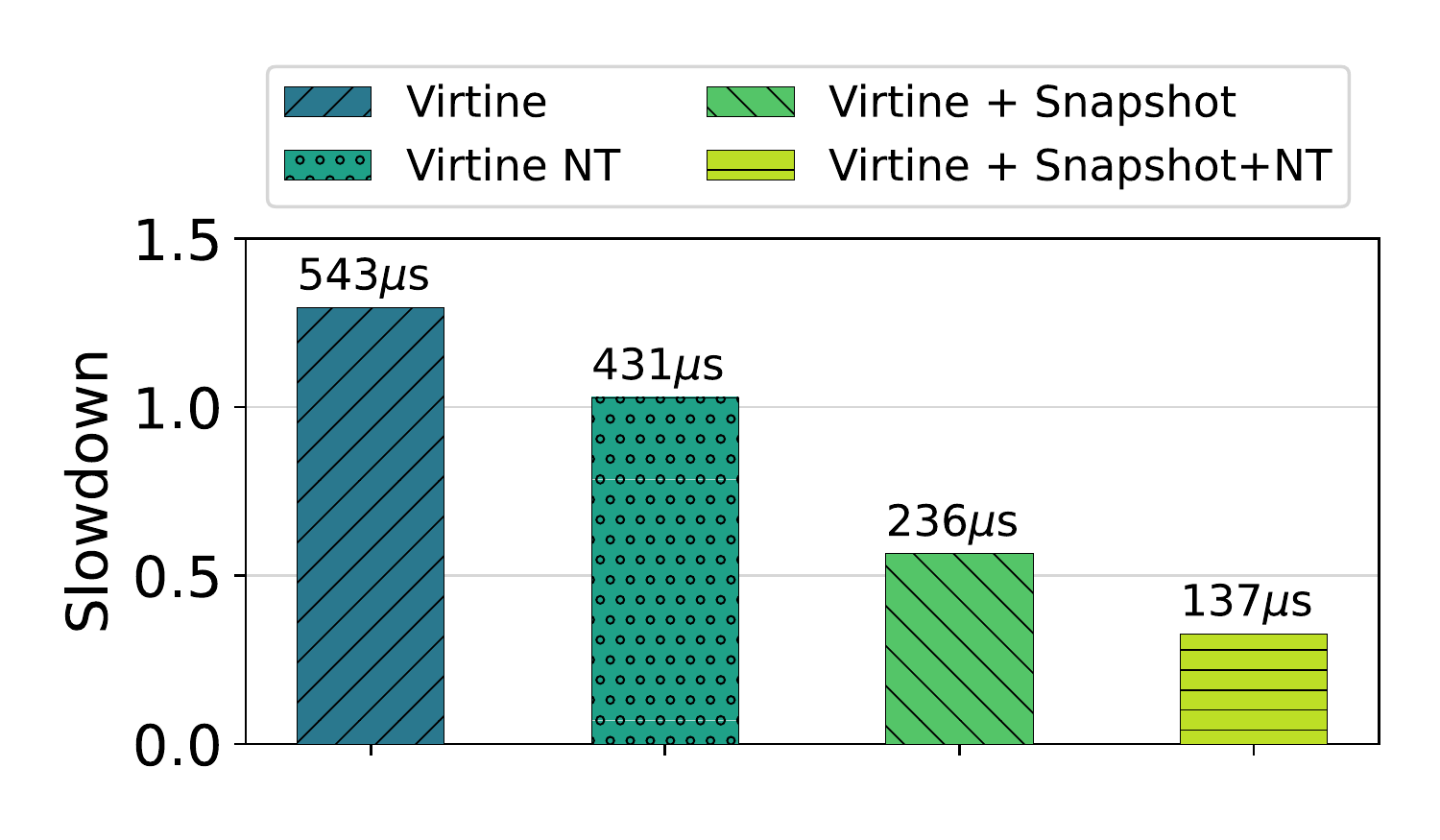}
    \caption{Slowdown of JavaScript virtines relative to native. The baseline latency is 419$\mu$s.}
    \label{fig:js}
\end{figure}

\paragraph*{Benchmarking and evaluation}
Figure~\ref{fig:js} shows the results of our Duktape implementation. The
virtine trial without snapshotting takes 125$\mu$s longer to execute than the
baseline. We attribute this to several sources, including the required virtine
provisioning and initialization overhead and the overhead to allocate and later
free the Duktape context. By giving programmers direct control over more
aspects of the execution environment, several optimizations can be made. For
example, snapshotting can be used as shown in
Figure~\ref{fig:snapshot-timeline} by including the initialization of the
JavaScript engine in the virtine's boot sequence. Doing so avoids many calls to
\verb|malloc| and other expensive functions while initializing. By taking
advantage of snapshotting in the case of the ``Virtine + Snapshot''
measurements, virtines can enjoy a significant reduction in overhead--roughly
$2\times$. Further, since all virtines are cleared and reset after execution,
paying the cost of tearing down the JavaScript engine can be avoided. By
applying both of these optimizations, the virtine can almost entirely avoid the
cost of allocating and freeing the Duktape context by retaining it--something
that cannot be done when executing in the client environment. Both of the
trials, ``Virtine NT'' and ``Virtine+Snapshot+NT'' are designed to take
advantage of this ``\textbf{\underline{N}}o \textbf{\underline{T}}eardown''
optimization in full. Note that the virtine is not executing code any {\em
faster} than native, but it is able to provide a significant reduction in
overhead by simply executing less code by applying optimizations. These
optimizations cause the overall latency to drop to 137$\mu$s, which effectively
constitutes the parsing and execution of the JavaScript code. Similar
optimizations are applied in SEUSS~\cite{CADDEN:2020:SEUSS}, which uses the
more complex V8 JavaScript engine, and thus avoids even more initialization
overhead.

\section{Discussion}
\label{sec:discuss}

In this section, we discuss how our results might translate to realistic
scenarios and more complex applications, the limitations of our current approach,
and other use cases we envision for virtines.

\subsection{Implications}
\label{sec:implications}

\paragraph*{Libraries}
In Section~\ref{sec:openssl}, we demonstrated that it requires little effort to
incorporate virtines into existing codebases that use sensitive or untrusted
library functions. In our example we assumed access to the library's code
(\texttt{libopenssl} in our case). While others make the same
assumption~\cite{GHOSN:2021:ENCLOSURES}, this is not an inherent limitation.
The virtine runtime could apply a combination of link-time wrapping and binary
rewriting to migrate library code automatically to run in virtine context.
Others have applied such techniques for software fault isolation
(SFI)~\cite{WAHBE:1993:SFI}, even in virtualized
settings~\cite{HALE:2014:GUARDMODS}.

\begin{figure}
    \centering
    \includegraphics[width=1.0\columnwidth]{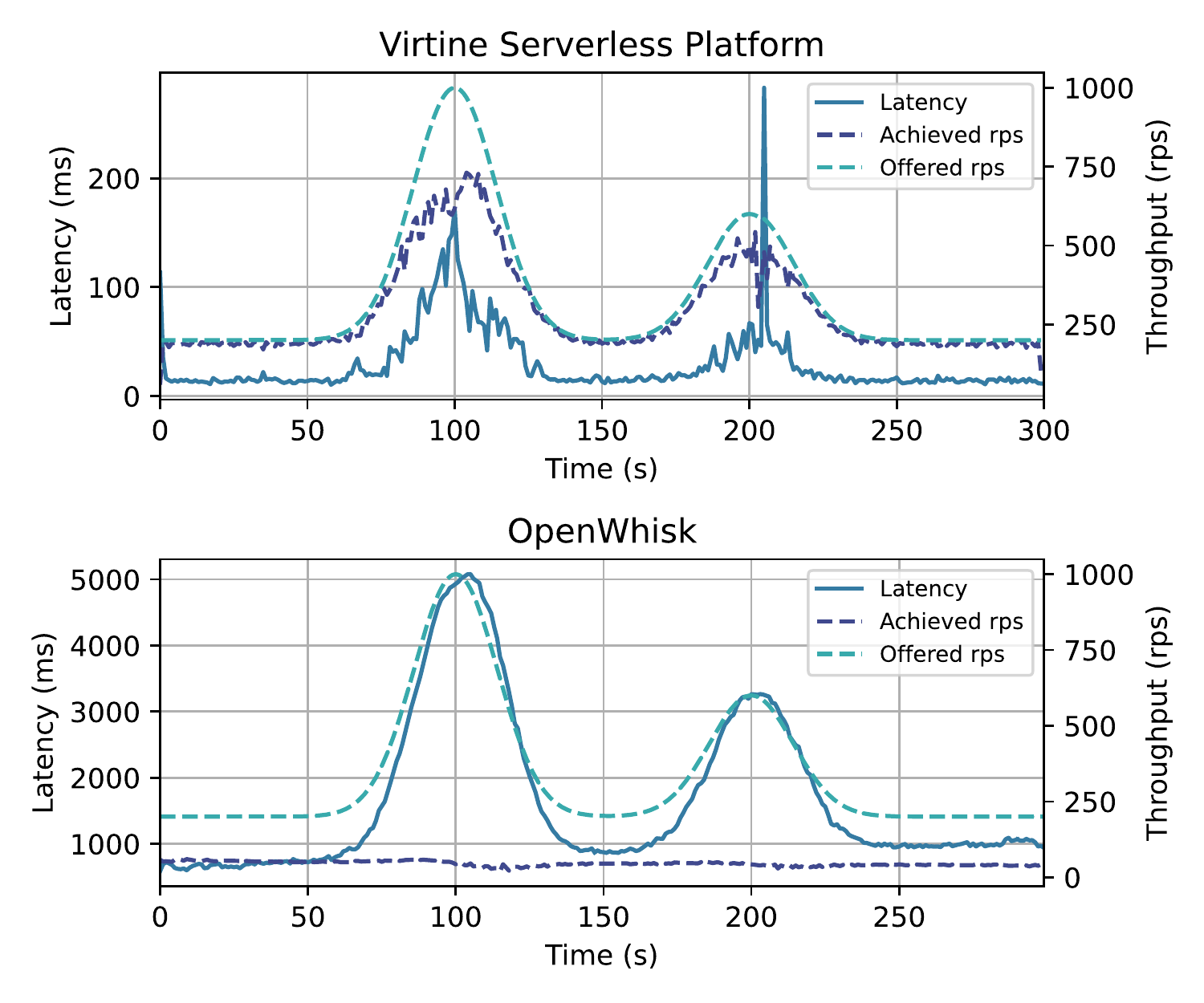}
    \caption{Serverless virtine performance compared to OpenWhisk's container-based platform.}
    \label{fig:virtine-serverless}
\end{figure}

\paragraph*{Serverless Functions}
While production serverless platforms and databases may use more complete
JavaScript engines like V8, we can reason about how the results of our Duktape
implementation would translate to these settings. Amazon Lambda, for example, constructs
a container to achieve the desired level of
isolation~\cite{AMAZON-LAMBDA-CONTAINER-REUSE:WEBSITE}. Here, the cost of
creating a process and allocating a V8 context is considerable.  If similar or
better isolation can be achieved with a virtine, then the cost of creating the
container can be eliminated. 

To determine the feasibility of serverless virtines, we implemented a prototype
serverless platform based on Apache's OpenWhisk framework~\cite{OPENWHISK} that integrates with our virtine
Duktape engine. The current
implementation only supports a single-node setup, and does not yet incorporate
an API gateway, a load balancer, or authentication mechanisms. In this platform,
which we call Vespid, users register JavaScript functions via a web
application, which produces requests to our framework's main endpoint. These requests are
handled by a concurrent server which runs each serverless function in a
distinct virtine (rather than a container) by leveraging the Wasp runtime API.
We measure the performance of invoking the same base64 JS encoding function
used in Section~\ref{sec:js} on our Vespid platform (which uses Duktape) and compare it to
vanilla OpenWhisk (which uses V8 via Node.js). The results are shown in Figure~\ref{fig:virtine-serverless}. We
measure end-to-end latency of both platforms, where client requests are
generated on the same node as the server to minimize differences in the
front-end implementations. We produce a series of concurrent function requests
(from multiple clients) against both platforms using
Locust~\cite{LOCUST:WEBSITE}, an off-the-shelf workload generator. This
invocation pattern involves an initial ramp-up period that leads to two bursts,
which then ramp down. Achieved throughput is shown on the dotted line. 
Vespid benefits from the lightweight virtine execution environment, Duktape's 
small image size, and Wasp's caching and snapshotting optimizations, leading
to low-latency responses for a bursty load pattern. However,
it is important to note that Vespid is a prototype that lacks many
features offered by OpenWhisk, including the high-performance V8 engine. Note
also that OpenWhisk's container engine does not employ optimizations such as 
container reuse and snapshotting seen in the
recent literature like SOCK~\cite{OAKES:2018:SOCK},
SEUSS~\cite{CADDEN:2020:SEUSS}, Faasm~\cite{SHILLAKER:2020:FAASM}, and
Catalyzer~\cite{DU:2020:CATALYZER}, which all provide cold-start latencies less
than 20ms. These results do show that a virtine-based serverless platform with
competitive performance is feasible. Serverless functions that leverage external
resources like S3 buckets can be facilitated with the appropriate virtine
client support via hypercall handlers. 

\paragraph*{User-defined Fucntions}
A similar model could be used to more strongly isolate UDFs from
one another in database systems. Postgres, for example, uses 
V8 mechanisms to isolate individual UDFs from one another~\cite{POSTGRESV8}, but they still execute in the
same address space. Because virtine address spaces are disjoint, they could help
with this limitation. Furthermore, virtines would allow functions in unsafe
languages (e.g., C, C++) to be safely used for UDFs.

\subsection{Limitations}
\label{sec:limitations}

\paragraph*{Workloads} While our current workloads represent components
that could be used in real settings, we do not currently integrate with
commodity serverless platforms or database engines. This integration is currently
underway, with OpenWhisk and PostgreSQL, respectively.

Applications that rely on high-performance
JavaScript will use a production engine like V8 or SpiderMonkey.
Our evaluation uses Duktape, which lacks 
features like JIT compilation, but has the benefits of compiling into a
small ($\sim$578KB) image and being easily portable.
We envision that with sufficient runtime support---in particular, a port of
the C++ standard library---a V8 virtine implementation is likely feasible. We
believe our evaluation demonstrates the potential for incorporating virtines with
high-level languages (HLLs).

\paragraph*{Language extensions} Currently, our C extension lacks the ability
to take advantage of functionality located in a different LLVM module (C source
file) than the one that contains the C function. Build systems used in C
applications produce intermediate object files that are linked into the final
executable. This means that virtines created using the C extension
are restricted to functionality in the same compilation unit. Solutions to this
problem typically involve modifying the build system to produce LLVM bitcode
and using whole program analysis to determine which functions
are available to the virtine, and which are not.


Automatically generated virtines face an ABI challenge for
argument passing. Because they do not share an address space with the
host, argument marshalling is necessary.  We leveraged LLVM to copy a
compile-time generated structure containing the argument values into the
virtine's address space at a known offset. Marshalling does incur an overhead
that varies with the argument types and sizes, as is typical with
``copy-restore'' semantics in RPC systems~\cite{BIRRELL:1983:RPC}. This affects
start-up latencies when launching virtines, as described in
\S\ref{sec:openssl}.

Virtines do not currently support nesting, but this is not an inherent
limitation. Virtines that dynamically allocate memory are possible with an
execution environment that provides heap allocation, but that memory is
currently limited to the virtine context. We believe secure channels to
communicate data between the virtine and host could be implemented with
appropriate hypercalls and library/language support. The virtine compiler could
identify and transform such allocation sites (e.g., malloc) using escape
analysis.

\paragraph*{Snapshotting performance} Wasp's snapshotting mechanism currently uses
\texttt{memcpy} to populate a virtine's memory image with the snapshot. This copying, as shown
in Figure~\ref{fig:image-size}, constitutes a considerable cost for a large virtine image.
We expect this cost to drop
when using copy-on-write mechanisms to reset a virtine, as in SEUSS~\cite{CADDEN:2020:SEUSS}.

\paragraph*{KVM performance} 
As we found in Section~\ref{sec:bootstrap-func}, KVM has performance penalties
due to its need to perform several ring transitions for each exit, and for VM start-up.  Some
of these costs are unavoidable because they maintain userspace control over
the VM. However, a Type-I VMM like Palacios or Xen~\cite{LANGE-PALACIOS-KITTEN-IPDPS10,
BARHAM:2003:XAV} can mitigate some software latencies incurred by virtines.

\paragraph*{Security} Our threat model makes assumptions that may not hold in
the real world. For example, a hardware bug in VT-x or a microarchitectural
side channel vulnerability (e.g., Meltdown~\cite{LIPP:2018:MELTDOWN}) could
feasibly be used to break our security guarantees.

\subsection{Other use cases}
In our examples, we used virtines to isolate certain annotated functions
from the rest of the program. This use case is not the only possible one. Below, we
outline several other potential use cases for virtines.

\paragraph*{Augmenting language runtimes}
We believe that HLLs present an \textit{incremental} path to using virtines,
i.e., the language runtime might abstract away the use of virtines entirely,
for example, to wrap function calls via the foreign function interface (Chisnall et al.
employed special-purpose hardware for this purpose~\cite{CHISNALL:2017:CHERIJNI}).
Virtines might also be used to apply security-in-depth to JIT compilers and
dynamic binary translators. For example, bugs that lead to vulnerabilities in
built-in functions or the JIT's slow-path handlers~\cite{ROHLF:2011:JIT} can be
mitigated by running them in virtine context (NoJITsu achieves this with
Intel's Memory Protection Keys~\cite{PARK:2020:NOJITSU}).  Polyglot
environments like GraalVM~\cite{WURTH:2013:GRAAL} could more safely use native
code by employing virtines.

\paragraph{Distributed services}
Because virtines implement an abstract machine model, are packaged with their
runtime environment, and employ similar semantics to
RPC~\cite{BIRRELL:1983:RPC}, they allow for location transparency. Virtines
could therefore be migrated to execute on remote machines just like containers,
e.g., for code offload. This could allow for implementing distributed services
with virtines, and for service migration based on high load scenarios,
especially when RPCs are fast, as in the datacenter~\cite{KALIA:2019:RPCS}. If
virtines require host services or hardware not present in the local machine,
they can be migrated to a machine that \textit{does}.


\section{Related Work}
\label{sec:related}

There is significant prior work on isolation of software components. However,
the received wisdom is that when using hardware virtualization, creating a new
isolated context for every isolation boundary crossing is too expensive. With
virtines, we have shown that, with sufficient optimization, these overheads can
be significantly reduced. Virtines enjoy several unique properties: they have an easy-to-use programming model,
they implement an abstract machine model that allows for customization of the execution environment and the hypervisor,
and because they create new contexts on every invocation, we can apply
snapshotting to optimize start-up costs. We now summarize key differences 
with prior work.

\paragraph*{Isolation techniques}
\label{sec:isolation}

The closest work to virtines is Enclosures~\cite{GHOSN:2021:ENCLOSURES}, which
allow for programmer-guided isolation by splitting libraries into their own
code, data, and configuration sections within the same binary. 
The security policy of Enclosures is defined in terms of \textit{packages}, but
with virtines, the security policy is defined and enforced at the level of
individual functions. While, like Enclosures, virtines can be used to isolate
library functions from their calling environment, they can \textit{also} be
used to selectively isolate functions from other users' virtines in a
multi-tenant cloud environment.

Hodor~\cite{HEDAYATI:2019:HODOR} also provides library isolation, particularly for high-performance
data-plane libraries. Gotee uses language-level
isolation like virtines, but builds on SGX enclaves rather than hardware
virtualization~\cite{GHOSN:2019:SECROUTINES}.

While TrustVisor~\cite{MCCUNE:2010:TRUSTVISOR} employs hardware virtualization to isolate application
components (and assumes a strong adversary
model), virtines enjoy a simpler programming
model.  SeCage uses static and dynamic analysis to automatically isolate
software components guided by the secrets those components
access~\cite{LIU:2015:SECAGE}. Virtines give programmers more control over
isolated components.  Glamdring also automatically partitions applications
based on annotations~\cite{LIND:2017:GLAMDRING}, but uses SGX Enclaves which
have more limited execution environments than virtines.

With Wedge~\cite{BITTAU:2008:WEDGE}, execution contexts (sthreads) are given
minimal permissions to resources (including memory) using default deny
semantics.  However, virtines are more flexible in that they need not use the
same host ABI and they do not require a modified host kernel. Dune is an
example of an unconventional use of a virtual execution environment that
provides high performance and direct access to hardware devices within a Linux
system~\cite{BELAY:2012:DUNE}. Unlike virtines, Dune's virtualization is at
{\em process} granularity. Similarly, SMV isolates multi-threaded
applications~\cite{HSU:2016:SMV}.

Several systems that support isolated execution leverage Intel's Memory
Protection Keys~\cite{INTEL_SYS_MAN} for memory
safety~\cite{SCHRAMMEL:2022:JENNY, VAHLDIEK:2019:ERIM, PARK:2020:NOJITSU,
DAUTENHAHN:2021:ENDOKERNEL, CONNOR:2020:PKU}. For virtines, we chose not to use
this mechanism since the number of protection domains (16) offered by the
hardware was insufficient for multi-tenant scenarios (e.g., serverless). Even
without this limitation, instructions that access the PKRU register would need
to be validated/removed, e.g., with binary rewriting, as is done in
ERIM~\cite{VAHLDIEK:2019:ERIM}. We leave the exploration of MPK and similar
fine-grained memory protection mechanisms for future work.

Lightweight-Contexts (LwCs) are isolated execution contexts {\em within} a
process~\cite{LITTON:2016:LWC}.  They share the same ABI as other contexts, but
essentially act as isolated co-routines.  Unlike LwCs, virtines can run an
arbitrary software stack, and gain the strong isolation benefits of hardware
virtualization. The Endokernel architecture~\cite{DAUTENHAHN:2021:ENDOKERNEL}
enables intra-process isolation with virtual privilege rings, but still maps
domains to the process abstraction, rather than functions.

Nooks~\cite{SWIFT:2002:NOOKS}, LXD~\cite{NARAYANAN:2019:LXD}, and Nested
Kernel~\cite{DAUTENHAHN:2015:NESTEDKERNEL} all implement isolation for kernel
modules.

Software Fault Isolation~\cite{WAHBE:1993:SFI} (SFI) enforces isolation by
instrumenting applications with enforcement checks at boundary crossings, and
thus does not leverage hardware support.

\paragraph*{Virtualization}
\label{sec:virt}
Wasp is similar in architecture to other minimal hypervisors (implementing
$\mu$VMs). Unlike Amazon's Firecracker~\cite{AGACHE:2020:FIRECRACKER-NSDI} or
Google's Cloud Hypervisor~\cite{GOOGLE-CLOUDVMM}, we do not intend to boot
a full Linux (or Windows) kernel, even with a simplified I/O device model. Wasp
bears more similarity to ukvm~\cite{WILLIAMS:2016:UKM} (especially the networking interface) and uhyve~\cite{LANKES:2017:UHYVE}.
Unlike those systems, we designed Wasp to use a set of pre-packaged runtime
environments.  We intend Wasp to be used
as a pluggable back-end (for applications, libraries, serverless platforms,
or language runtimes) rather than as a stand-alone VMM.

\paragraph*{Serverless}
\label{sec:serverless}
Faasm~\cite{SHILLAKER:2020:FAASM} employs SFI for
isolated, stateful serverless functions, partly based on the premise that
hardware virtualization is simply too expensive. We show that this is not
necessarily the case. Cloudflare uses V8 JavaScript ``isolates'' to reduce VM cold-start at the language level~\cite{CLOUDFLARE-WORKERS}.
Others have developed optimized serverless systems based on the observation
that a significant fraction of start-up costs can be attributed to language
runtime and system software initialization, a task often duplicated across
function invocations~\cite{OAKES:2018:SOCK, CADDEN:2020:SEUSS,
DU:2020:CATALYZER}. These systems achieve start-up latencies in the
sub-millisecond range with aggressive caching of runtime state, which we also employ.

\paragraph*{Execution environments}
\label{sec:exec}
Jitsu~\cite{MADHAVAPEDDY:2015:JITSU} allows Unikernels to be spawned on demand
in response to network events, but does not allow programmers to invoke
virtualized environments at the function call granularity.
There is a rich history of combining language and OS research. 
Typified by MirageOS~\cite{MADHAVAPEDDY:2013:UNIKERNELS}, writing kernel components in a high-level language
gives the kernel developer more flexibility in moving away from legacy interfaces. 
It can also shift the burden of protection and isolation~\cite{PORTER:2011:DRAWBRIDGE,
HUNT:2007:SINGULARITY}. 

\section{Conclusions and Future Work}
\label{sec:conc}

In this work, we explored the design and implementation of {\em
virtines}---light-weight, isolated, virtualized functions, which can provide
fine-grained execution without much of the overheads of traditional hardware
virtualized execution environments. We probed the lower limits of hardware
virtualization and presented Wasp, an embeddable hypervisor designed
for virtines with microsecond start-up latency and limited slow-down. Wasp
allows programs to easily create isolated contexts with tunable isolation
policies. We introduced a compiler extension that allows developers to use
virtines with simple code annotations, and explored how they can be used to ease
virtine deployment. We demonstrated that integrating virtines with existing
library code takes little effort. We modified an off-the-shelf
JavaScript engine to use virtines, and explored how high-level
languages can take advantage of them. In future work, we plan to 
explore other virtine applications in HLLs, for example in JIT engines. We also plan to
investigate automatic virtine environment synthesis.

\begin{acks}
	We thank the anonymous reviewers and our shepherd, Martin Maas. We also thank
	the anonymous reviewers from OSDI '20 and EuroSys '21 for their encouraging
	and constructive feedback that helped improve the paper significantly. We
	thank Andrew Chien, Tyler Caraza-Harter, Boris Glavic, and Peter Dinda for
	enlightening discussions and suggestions. We also thank Andrew Neth, Cooper
	Van Kampen, and Griffin Dube for their help field testing the artifact
	scripts on a variety of machines. This work was made possible with support
	from the \grantsponsor{01}{United States National Science Foundation
	(NSF)}{https://nsf.gov} via grants \grantnum{01}{CNS-1718252},
	~\grantnum{01}{CNS-1730689}, ~\grantnum{01}{REU-1757964},
	and~\grantnum{01}{CNS-1763612}.
\end{acks}


\balance
\bibliographystyle{ACM-Reference-Format}
\bibliography{kyle}

\newpage
\appendix
\section{Artifact Appendix} 

\subsection{Abstract}

This artifact comprises our Wasp $\mu$hypervisor, our Clang compiler
extensions, our LLVM pass, the experimental code, scripts, and data used
for the paper, as well as benchmarks and example code. Users should be able to re-create our experiments and compare
results on a broad range of hardware. In our Github
repo\footnote{\url{https://github.com/virtines/wasp}} we have detailed
instructions on how to run Wasp, and compile code with virtine support.

\subsection{Description \& Requirements}

\subsubsection{How to access}

All code for virtines and Wasp is organized under the Virtines Github
organization. This can be accessed at \url{https://github.com/virtines}. The
primary repository that will be used for this Artifact is Wasp, which is at
\url{https://github.com/virtines/wasp}. The DOI for our artifact is 10.5281/zenodo.6350453.

\subsubsection{Hardware dependencies}

Wasp currently supports x64 hardware, including AMD and Intel. Hardware virtualization extensions are 
a requirement (SVM on AMD, VT-x on Intel), so hardware from the last decade or so should work fine. 

\subsubsection{Software dependencies}\
The primary dependence is on a hosted hypervisor framework (namely, kvm on
Linux). While early versions of Wasp ran on Hyper-V, the port of our compiler
extensions and benchmarks is not yet complete, so for this artifact we focus on
Linux. We also rely on the following packages and libraries:

\begin{itemize}
    \item netwide assembler (\texttt{nasm})
    \item \texttt{curl} development libraries (\texttt{libcurl-dev} on Debian-based distros)
    \item Clang C compiler (version 10 or newer)
    \item LLVM and development headers (\texttt{llvm} and \texttt{llvm-dev})
    \item \texttt{cmake}
    \item Virtual environments for Python3 (\texttt{python3.8-venv})
\end{itemize}

It is possible to run Wasp in a nested virtualization environment (as long as
nested virtualization with kvm is enabled), but we recommend running on a bare-metal
machine to get reasonable performance. More details on the build prerequisites can be found in the
instructions in the repo (\texttt{README.md}).

\subsubsection{Benchmarks} 
All relevant benchmarks are included in the repo. Experimental results are reproduced with these benchmarks as described below.

\subsection{Set-up}

We recommend evaluators follow the guidance in our repo (\texttt{README.md}).
To build Wasp, see the sections titled ``Environment Setup,''
``Prerequisites,'' and ``Building and Installing.'' As described there, we
recommend an environment like Cloudlab or Chameleon Cloud. We provide
a Cloudlab profile and instructions for Chameleon in our README document. 


\subsection{Evaluation workflow}

Once Wasp is set up, reproduction of experimental results is a mostly automated
process, though evaluators are free to focus on individual experiments.
Before that, to ensure that Wasp is functional once built, users can run
\texttt{make smoketest} to ensure everything is working properly. See our
README ("Running Virtine Tests") for more detail. Once Wasp is functional,
experimental results can be easily generated. 

\subsubsection{Major Claims}

\begin{itemize}

    \item \textit{(C1): The core components of virtual context creation comprise only a few tens of thousands of cycles. We show this in experiment (E1) described in Section 4.2, whose results are shown in Table 1.}

    \item \textit{(C2): The latency to run a function in different processor modes can vary (e.g., 16-bit mode is cheaper on some microarchitectures), presenting an opportunity for optimization when building virtual contexts. We show this in experiment (E2) in Section 4.2 of the paper, whose results are depicted in Figure 3. }

    \item \textit{(C3): A runtime system that boots a basic server in a minimal execution environment can achieve 
        response times $<$1ms, even without optimizations. We show this in experiment (E3) in Section 4.2 of the paper, whose results
        are shown in Figure 4.}

    \item \textit{(C4): Virtual context creation latencies with Wasp approach the hardware limit of the vmrun/vmcall instruction
        by employing optimizations. We show this in experiment (E4) in Section 5.2 of the paper (Figure 8). Note that we see
        Figure 2 as a subset of the results in Figure 8, so we elide it in the artifact. }

    \item \textit{(C5): Virtine creation overheads can be amortized with roughly 100$\mu$s of work. In finer-grained scenarios,
        snapshotting can reduce overheads significantly, pushing the amortization point down by about 10$\times$. We show this
        in experiment (E5) in Section 6.1 of the paper (Figure 11). }

    \item \textit{(C6): Once virtine image size reaches around 2MB, start-up latency becomes bottlenecked by memory
        bandwidth. We show this in experiment (E6) in Section 6.2 of the paper (Figure 12). }

    \item \textit{(C7): An HTTP server using our virtine compiler extensions experiences a less than 20\% drop in throughput
        relative to a native environment. We show this in experiment (E7) in Section 6.3 of the paper (Figure 13). }

    \item \textit{(C8): Virtines can be integrated with an off-the-shelf Javascript engine, with acceptable ($<1.5\times$)
        slow-downs ($\sim$2$\times$). Snapshotting improves performance when environment setup in the virtual context is non-trival. We show this in experiment (E8) in Section 6.5 of the paper (Figure 14).}

\end{itemize}

\subsubsection{Experiments}

To re-run all experiments, you can simply run \texttt{make artifacts.tar} as described in our README. This will take roughly five minutes
to run to completion and generate data and plots, which can then be found in \texttt{artifacts.tar}. Once the results are generated,
they can be compared with the data from the paper (which can be also be found in \texttt{data\_example/gold/}). We have also provided data
we have generated on many other machines. These can also be found in \texttt{data\_example/*}; the machine and software environment
descriptions can be found in\\
\texttt{data\_example/README.md}. 

We outline individual experiments below.\\

\textit{Experiment (E1): [Boot Breakdown] [1 sec.]: This experiment evaluates the various components that comprise booting 
a virtual context. Use this to evaluate claim (C1). }

\textit{[How to]}
Run \texttt{make figure1\_data}. 

\textit{[Results]}
The results will appear in\\
\texttt{data/figure1\_data.csv}. You should see that
the total average cycle counts less than $\sim$100K cycles (ignoring the first
run). The transition to protected mode (\texttt{prot}) and the identity mapping
(\texttt{id map}) will be the most expensive components of the boot process. \\

\textit{Experiment (E2): [Mode latency] [5 sec.]: This experiment evaluates the time to run a recursive implementation
of fib(20) in 16-bit mode, 32-bit (protected) mode, and 64-bit (long) mode. Use this to evaluate claim (C2).}

\textit{[How to]} 
Run \texttt{make fig3.pdf}. 

\textit{[Results]}
You can see the results in \texttt{fig3.pdf}. In most machines the time to run the function will vary with processor modes,
but on some there is little difference. The point here is that there is room for a virtine compiler to leverage hardware knowledge
to optimize code generated for virtine context. \\

\textit{Experiment (E3): [Echo server] [5 sec.]: This experiment shows that with a minimal virtual exection context, an
HTTP server can achieve response times $<$1ms. Use this to evaluate claim (C3).}

\textit{[How to]} 
Run \texttt{make fig4.pdf}. 

\textit{[Results]}
You can see the results in \texttt{fig4.pdf}. You should see that the time to get an HTTP response should be somewhere between
100K and 500K cycles. \\

\textit{Experiment (E4): [Context creation] [5 sec.]: This experiment shows that Wasp can achieve start-up latencies
close to the hardware limit (the vmrun/vmcall instruction on x86 hardware). Use this to evaluate claim (C4). }

\textit{[How to]} 
Run \texttt{make fig8.pdf}. 

\textit{[Results]}
You can see the results in \texttt{fig8.pdf}. You should see that the
``Wasp+C'' and ``Wasp+CA'' bars appear relatively close to the vmrun bar and
outperform pthreads. \\

\textit{Experiment (E5): [Virtine overheads] [30 sec.]: This experiment demonstrate how much computation
is necessary to amortize virtine creation overheads. Use this to evaluate claim (C5).}

\textit{[How to]} 
Run \texttt{make fig11.pdf}. 

\textit{[Results]}
You can see the results in \texttt{fig11.pdf}. You should see the bars even out in the 100$\mu$s range,
indicating that 100$\mu$s of computation are necessary to amortize creation overheads. Snapshotting should
significantly improve call latency for smaller image sizes. \\

\textit{Experiment (E6): [Image size impact] [3 sec.]: This experiment demonstrates the impact
of image size on virtine creation latency. Use this to evaluate claim (C6).} 

\textit{[How to]} 
Run \texttt{make fig12.pdf}. 

\textit{[Results]}
You can see the results in \texttt{fig12.pdf}. You should see a knee in the
curve somewhere around 1-2MB. This is where virtine creation is becoming memory
bandwidth bound. Where exactly the knee occurs depends on the memory copy
bandwidth of the machine. \\

\textit{Experiment (E7): [HTTP server] [1 min.]: This shows the latency and throughput of serving
HTTP requests in a virtine (with and without snapshotting optimization) vs. native execution. 
Use this to evaluate claim (C7).} 

\textit{[How to]} 
Run \texttt{make fig13\_lat.pdf} then\\
\texttt{make fig13\_tput.pdf}. 

\textit{[Results]}
You can see the results in \texttt{fig13\_lat.pdf} and \texttt{fig13\_tput.pdf}. Expect to see a little more
than 2$\times$ increase in latency and 2$\times$ drop in throughput relative to native. Snapshotting
may actually reduce performance on this experiment on machines with limited memory bandwidth. Most of
the performance drop is caused by hypercall interactions. \\

\textit{Experiment (E8): [Javascript virtines] [3 sec.]: This shows the slowdown of 
launching Javascript virtines with various optimizations. Use this to evaluate claim (C8).}

\textit{[How to]} 
Run \texttt{make fig14.pdf}. 

\textit{[Results]}
You can see the results in \texttt{fig14.pdf}. You should see that snapshotting has a real effect
given the amount of runtime initialization that takes place in the Duktape JavaScript engine. The slowdown
for virtines without optimizations (the leftmost bar) should be in the 1.5--2$\times$ range. 


\subsection{Notes on Reusability}
\label{sec:reuse}
If you would like to explore using embedded Wasp with your programs, we provide
additional guidance in the ``Embedding Wasp'' section of our repo's README.
Developers can interact with the runtime library directly using our API or
indirectly using our compiler extensions. 

\balance

\end{document}